\documentclass[aps,pre,twocolumn,showpacs,epsfig,epsf,amssymb,floatfix]{revtex4-1}
\usepackage{graphicx}
\usepackage{amssymb}
\usepackage{amsmath}
\usepackage{subfigure}
\usepackage{booktabs}
\usepackage{float}
\usepackage{longtable}
\begin{document}
\newcommand{\beginsupplement}{%
        \setcounter{table}{0}
        \renewcommand{\thetable}{S\arabic{table}}%
        \setcounter{figure}{0}
        \renewcommand{\thefigure}{S\arabic{figure}}%
     }

%\widetext

\title{Bacterial chromosome organization II: few special cross-links, cell confinement, and molecular crowders play the pivotal roles.}
\author{Tejal Agarwal$^{1}$ } \email{tejal.agarwal@students.iiserpune.ac.in}
\author{G.P. Manjunath $^2$}
\author{Farhat Habib$^3$}
\author{Apratim Chatterji$^{1,4}$} 
\email{apratim@iiserpune.ac.in}
\affiliation{
$^1$ IISER-Pune, Dr. Homi Bhabha Road,  Pune-411008, India.\\
$2$ Department of Biochemistry and Molecular Pharmacology, NYU Langone Medical Center, New York, NY 10016, USA. \\
$3$ Inmobi - Cessna Business Park, Outer Ring Road, Bangalore-560103, India. \\
$^4$ Center for Energy Science, IISER-Pune,  Dr. Homi Bhabha Road,  Pune-411008, India.
}

\date{\today}
\begin{abstract}
Using a bead-spring model of bacterial DNA polymers of {\em C. crescentus} and {\em E. coli}
we show that just $33$ and $38$ effective cross-links at special positions along the chain contour of the DNA can lead to the large-scale organization of the DNA polymer, where confinement effects of the cell walls play a key role in the organization. The positions of the $33$ cross-links along the chain contour are chosen from the contact map data of {\em C. crescentus}. We represent  $1000$ base pairs as a coarse-grained monomer in our bead-spring flexible ring polymer model of the DNA. Thus a $4017$ beads on a flexible ring polymer represents the {\em C. crescentus}  DNA with $4017$ kilo-base pairs. Choosing suitable parameters from our preceding study, we also incorporate the role of molecular crowders and the ability of the chain to release topological constraints. We validate our prediction of the organization of the {\em C. crescentus} with available experimental contact map data and also give a prediction of the approximate positions of different segments within the cell in 3D. For the  {\em  E. coli} chromosome with $4.6$ million base pairs, we need around $38$ effective cross-links with cylindrical  confinement to organize the chromosome. We also predict the 3D organization of the {\em  E. coli} chromosome segments within the cylinder which represents the cell wall.   
\end{abstract}
\keywords{}
%\pacs{87.15.ak,82.35.Pq,82.35.Lr}

\maketitle
\section{Introduction}
In the last few decades there has been a lot of interest to understand the emergent organization of the chromosomes at micron length scales starting from length scales of $30 nm$ fiber or higher \cite{le,jun,aiden,tjong,joyeux,marenduzzo,brocken,kremer,cagliero}.
%in the bacteria or higher organisms  %The organization of the chromosome helps in all the cellular processes from gene expression to DNA replication. The experiments like FISH, live imaging of the chromosome, Hi-C and its variants (CCC, 5C) as well as the polymer physics models have given deep insights into the problem. But we are far from the complete understanding of the organization of the chromosome at large length scales. In-vivo various factors may play a role in the organization of the chromosome. Some of the factors which have gained wide interests in the organization of the DNA are: DNA binding proteins, confinement by the nucleus in higher organisms or by cell in bacteria, super-coiling of the circular chromosome in bacteria, activity of enzyme topoisomerase to release topological constraints by allowing the chain to cross itself. 
In our previous papers \cite{epl,jpcm}, we studied models of the bacterial DNA-polymer  with cross-links (CLs) at very specific positions, but without considering confinements effects  due to the cell wall. We chose the specific monomer pairs which constitute the cross-links from the Hi-C contact map of bacteria {\em C. crescentus} and {\em E. coli} \cite{cagliero,le}. We showed that very few cross-links (less than $3\%$ of the chain monomers) could give an average structure to the DNA polymer. The CLs may mimic the effect of DNA-binding proteins which bind two different specific segments of the DNA chain together.  

In the paper just preceding this, we reported the important roles that molecular crowders and the release of topological constraints play in the organization of the model chromosome of bacteria {\em C. crescentus} (and  {\em E. coli}) with CLs at specific positions along the chain \cite{paper3}. Using Monte-Carlo (MC) simulations of the polymer with cross-links (CLs) at very specific positions we showed that the release of topological constraints  is essential for the polymer to reach a unique and specific structure. We controlled the relative ease with which a chain could release the topological constraints by taking different values of monomer diameter.  We also showed that the crowding environment of the bacterial chromosome could also help in the organization of the polymer. We did not model the effect of crowding environment by taking additional particles as done in several previous studies \cite{ralf,elcock,kojima}, but introduce the weak attraction ($=0.3k_BT$, the thermal energy) between the monomers of the model DNA-polymer to model the role of crowders. The small attraction between the monomers of the polymer mimics the role of molecular crowders as it has been reported in the previous studies that the crowding environment in the cytoplasm of the cell increases the effective attraction between the different DNA segments \cite{ralf,elcock,joyeux}.  

There are many previous studies which focus on the organization and dynamics of the polymer under the different shapes of confinement \cite{jun,wu,axel,jung,hongsuk,le,william,debashish,byha,austin,tung,wang,dai,milchev}. It is argued in several studies that the confinement plays a crucial role in the organization of the chromosomes in bacteria as well as in the higher organisms \cite{jun,axel}. In the higher organisms the chromosomes are confined by the spherically shaped nucleus membrane and in rod shape bacteria chromosomes are confined within the capsule shape cell membrane.  
In a recent study of polymers under spherical confinement it has been shown that the confinement slows down the dynamics of the polymers (glassy-dynamics) and the glassy dynamics helps in the segregation of the chromosomes in the nucleus for the human chromosomes \cite{hongsuk}. Other studies on bacterial chromosomes organization took the cylindrical confinement of the cell into account and showed that the chromosome is arranged in a bottle-brush like structure with several loops emanating from the central backbone, and the backbone is arranged parallel to the length of the cylinder \cite{le,william}. In another separate study of bottle-brush polymer (backbone and side-loops), it is reported that the backbone attains the helical structure spontaneously when confined to the small cylindrical volume and this structure is maintained for the different aspect ratio of the cylinder for high packing fractions \cite{debashish}. Several researchers have also reported the dynamics of the polymer under cylindrical confinement for various aspect ratios \cite{jun,byha,ralf}. In these papers, it is pointed that in the presence of cylindrical confinement the segregation of the two ring polymers is entropically driven. Keeping such studies in mind, we expect that confinement 
constraint could play a pivotal role in modifying the 3D organization of the DNA-polymers of {\em C. crescentus} and {\em E. coli}. The presence of
cylindrical cell walls can modify the spherical globule organization of DNA-polymer that we have obtained in our previous studies. 

Hence, in this paper we focus on investigating the role of cylindrical confinement of the cell wall on the organization of the DNA polymer, in addition to the contributions  from (a) DNA-binding proteins at specific positions (modeled by specific CLs from experimental contact map), (b) molecular crowders (modeled by weak attraction between the monomers) and (c) the release of topological constraints (modeled by small monomer bead diameter). We assume that the shape of the cell to be a cylinder and neglect the effect of spheroidal end caps. As before, we use equilibrium statistical mechanics (Monte Carlo simulations: Metropolis algorithm with the Boltzmann distribution of energy) and bead-spring model of a flexible polymer to study DNA-organization, though we are aware of the fact that a living cell is a non-equilibrium driven system. The assumption of a flexible polymer is tenable, as we are looking at coarse-grained picture of the chromosome, and our one monomer represents $1000$ base pairs. In the present investigation, we give detailed step by step  calculations and develop understanding of the emergence of organization of DNA of {\em C. crescentus} in the presence of confinement. At the end, we give our prediction of the organization of  {\em C. crescentus} using nearly half the number of CLs than we what used in \cite{epl}.  We also now have a better match of predicted simulation contact-map with the coarse-grained experimental contact map that  we had previously used  as a validation of our prediction in \cite{epl}. 

We use the same simulation protocols to predict the organization of {\em E. coli}, where we also point a crucial difference in the  protocol we followed. The difference in the protocol to obtain the results for {\em E. coli} lies primarily due to the fact that for {\em C. crescentus} the {\em ori} is observed to remain tethered at one end of the cylinder, whereas there is no such constraint for the {\em ori} of bacteria {\em E. coli}. The predicted 3D organization presented in this paper is consistent with our previous prediction of the 2D organization of {\em E. coli} (without confinement) given in \cite{jpcm,epl}.

The organization of the manuscript is as follows:  we first describe our model and simulation methods for the polymer with the various number of CLs under cylindrical confinement, where we primarily discuss details of modeling confinement. The other aspects of the modeling are very similar to what is described in the preceding paper; here we briefly mention some aspects for the sake of completeness. Next, we present our results in the Results sections.  In the end, we give our prediction of the organization for the chromosomes of the $2$ bacteria, and finally, conclude with the discussion section.

\section{Model and methods}
We model the circular chromosome of bacteria {\em C. crescentus} and {\em E. coli} as a coarse-grained bead-spring ring polymer ($4017$ and $4642$ monomers) with the specific cross-links, whose positions along the chain contour are chosen from the experimental contact maps as previously. For the detail refer \cite{epl,jpcm,paper3}. The nearest neighboring monomers along the chain interact by the harmonic potential of spring constant $\kappa=200k_BT/a^2$, where $a$ is the bond length; and  $a=1$ and $k_BT=1$ set the length and time scales for our simulations. We also model the excluded volume interaction between the monomers by suitably shifted Lennard Jones potential with a cutoff at $r_c=2^{1/6}\sigma$. Here $\sigma$ corresponds to the bead diameter, and we have set the value of $\sigma=0.2a$ to allow the chain crossings. In the preceding paper, we have shown that for $\sigma=0.2a$ the equilibrium organization of the ring polymer with CLs starting from very different initial conditions remains the same, within statistical fluctuations \cite{paper3}: the calculation of the Pearson correlations to compare the equilibrium organization of the polymers from  MC runs starting from independent initial conditions give high values close to $1$. The monomers which constitute the CLs are held together at a distance of $a$ by the harmonic potential with spring constant $\kappa_c=200k_BT/a^2$. 

To fix the degree and shape of confinement in our model, we take the help of previous experimental observations. It is known that the bacteria {\em C. crescentus} has a capsule-like shape.  We approximate the capsule shape  to be a cylinder for the sake of simplicity, and the DNA polymer is confined within a cylinder of aspect ratio (diameter: length) $\approx 1:7.5$ with planar ends. We intentionally chose a longer aspect ratio than the expected $1:5$ to check whether the final organization ends up in a configuration with a ratio closer to $1:5$; and indeed it does. Moreover, it has already been suggested previously \cite{le}  that the {\em C. crescentus} DNA could have a bottle brush structure with plectonemes (super-coiled segments of the DNA) emanating out in different directions 
from a central backbone. It is strongly believed that plectonemes play a significant role in DNA organization  at large length scales \cite{le,joyeux}. Though we do not explicitly consider plectonemes in our model,  the CLs that we use from the experimental contact maps may incorporate the effect of plectonemes.  This is because the DNA segments within plectonemes which are in spatial proximity in the 3D space will be reflected in the experimental contact maps.  Experimentally, it is known that  the average length of the plectoneme segment is $10-15$ kilo-BP \cite{le} which is up to $15$ monomers in our model. Since a plectoneme is coiled on itself, the maximum spatial length  of a plectoneme can be $\sim 7a$ in our model units. We take $7a$ to be the radius of  the cylinder; diameter $D=14a$ allows two plectonemes  to be  radially opposite to each other in the cylinder. The length of our confining cylinder was then fixed 
at $108a$ to maintain the given aspect ratio. 

In our previous works, we started our Monte-Carlo simulations of polymer with CLs from specially designed $9$ independent initial conditions without any confinement.  But in the presence of confinement, we cannot start the simulation with the same initial conditions as we used before, as those initial configuration will violate the constraint of the confinement. Further, if we start our simulation with the random initial configuration of the ring polymer and with the BC-2 set of CLs with $153$ CLs, the polymer forms a blob at the center of the cylinder, and the polymer will be unable to relax. Hence, to systematically investigate the role of confinement, we first start with a ring polymer (without CLs) in the cylinder and then observe how entropy makes the polymer spread out. We have 12 different initial conditions, described later. Further, it is known for bacteria {\em C. crescentus} that the DNA segment where the replication starts (origin of replication termed as {\em ori})  is attached to one end of the cylinder.  We incorporate  this constraint at this stage by attaching the  monomer which we have given the index $1$, to one pole of  the cylinder (see Fig. \ref{fig111}). The positions of the first monomer remain unchanged throughout the simulation. We first observe the organization of the ring polymer (without CLs) which is confined within a cylinder with the first monomer tethered at one pole of the cylinder.

In the second part of our study, we introduce the CLs at specific positions to this spread-out configuration of the polymer, allow the system to relax and reach equilibrium and carry out further analysis at the second stage of our computations.The effect of CLs are added in the model by introducing a quadratic spring potential between a specific pair of monomers with very low strength, and then slowly increase the strength as we allow the system to relax. At the end of this step, even if two cross-linked monomers are far away in space in the initial starting configuration, they come close to each other at the end of equilibration and thereafter maintain a distance of $a$. The cross-linked monomers drag the adjacent segments along with them. However, confinement  already restricts the possible polymer configurations in the 3D space. Hence instead of $153$ CLs (BC-2 CL-set in \cite{epl}) for bacteria {\em C. crescentus}, we just add $60$ and $49$ number of CLs at specific locations along the chain; we refer these as $BC^{'}$ and BC-1 set of CLs.  The list of CLs corresponding to the $BC^{'}$ and $BC-1$ CL set for the bacterial chromosome of {\em C. crescentus} (and {\em E. coli}) are given in the table S1 of supplementary section of this paper. We choose the number of CLs as per our experience gained with our previous systematic studies, where we checked for the minimal number of CLs required to organize the polymer into a particular structure. The minimal number of CLs which were required to obtain organization in the DNA ring polymers without confinement were $159$ and $153$ CLs for {\em E. coli} and {\em C. crescentus}, respectively \cite{epl,jpcm}. These correspond to $82$ and $60$ {\em effective} CLs, respectively, and we refer to that set as BC-2 set of CLs. 

For the present study, the $BC^{'}$ set of CLs are chosen by setting a suitably high frequency-cutoff (of the two segments to be in found in the spatial proximity) in the contact map. These $60, 49$ CLs correspond to $33$ and $26$ {\em effective} CLs \cite{effectivecl} and we refer to them as the $BC^{'}$ and BC-1 set of CLs for {\em C. crscentus}, respectively. The $49$ CLs in the BC-1 CL set are same as the  CLs which were taken in our previous work in the absence of confinement \cite{epl}, and then we did not observe any large-scale organization. Also, note that one CL-set is the subset of the other, i.e., $BC^{'}$ CL set has all the cross-links that are present in BC-1 set in addition to some extra CLs. Also, for our studies with {\em E. coli}, we have used $77$ CLs (which is $38$ {\em effective} CLs) to obtain the 3d-organization of the chromosome. Note again that the $77$ CLs are significantly lesser compared to the $159$ CLs which we use in our study of bacteria {\em E. coli} \cite{jpcm}. The details of the equilibration  and the design of initial conditions are given in the Results section.

In the third and last part of our study of DNA-polymer with confinement,  we added  a weak Lennard Jones attraction acting between all the monomers and compared the organization in the presence and absence of attraction. The attraction between the monomers mimics the effect of molecular crowding. In the absence of attraction, we just have the WCA (Weeks Chandler Anderson) potential acting between the monomers, whereas, we model the weak attraction between monomers using the Lennard Jones potential ($V=4\epsilon\left[(\sigma/r)^{12}-(\sigma/r)^6\right]$) with the  cutoff of the potential at $3\sigma$. Here, the parameter $\epsilon$ determine the strength of attraction between the monomers which we chose as $\epsilon = 0.3 k_BT$. Our choice of the value of attraction strength parameter $\epsilon$ is based on our study of the organization of the DNA ring-polymer with CLs for different values of the parameter $\epsilon$, but without incorporating the confinement effects in \cite{paper3}. We found that for the value of the parameter $\epsilon=0.3k_BT$ the positional correlation colormaps of different segments of the polymer from the $9$ independent runs match with each other with a very high value of Pearson correlation coefficient.

\section{Results}
  
\begin{figure}[!hbt]
\includegraphics[width=0.49\columnwidth]{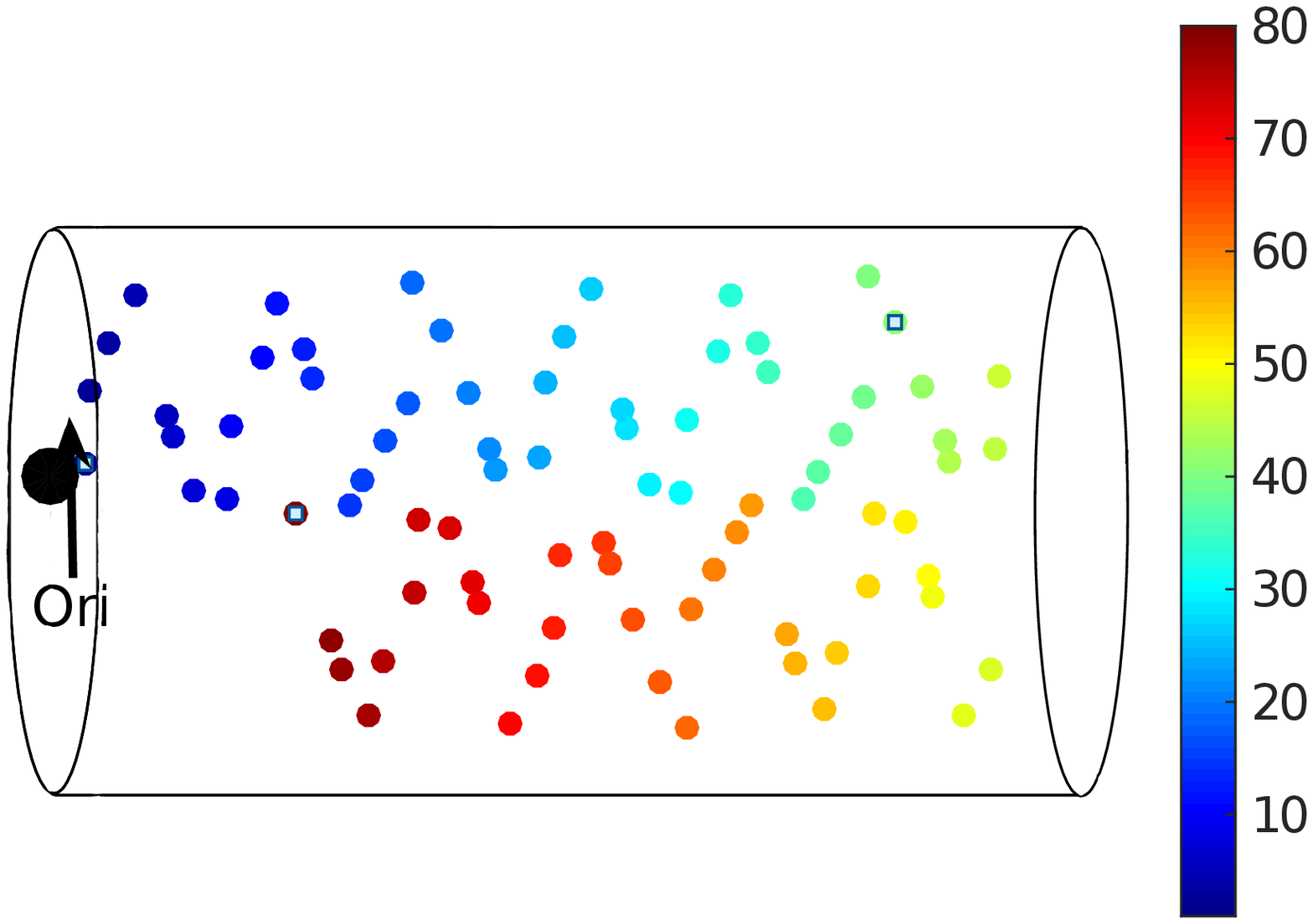} 
\includegraphics[width=0.49\columnwidth]{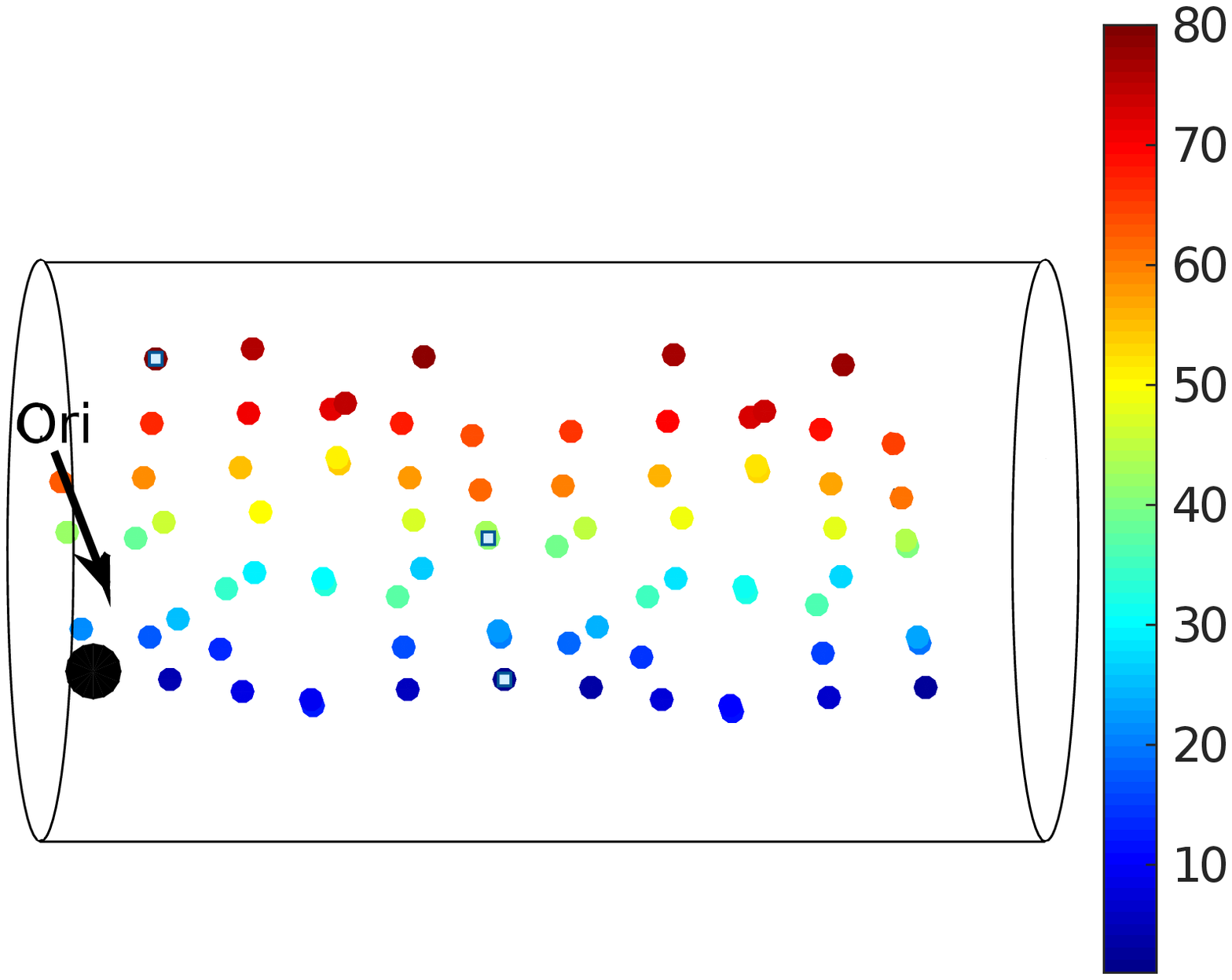}
\caption{\label{fig111}
The figure shows the $2$ different initial arrangements of the polymer inside the cylinder. The monomers are represented by the color according to their index
in the contour from blue to red. For better visualization we have only plotted the positions of every 50$^{th}$ monomer. Thus the index 80 in the colorbar corresponds to the monomer numbered $4000$. The first monomer 
of the chain is colored black and marked as ori. 
}
\end{figure}

We study the effect of cylindrical confinement in the organization of DNA ring polymer for the bacteria {\em C. crescentus}.
For this, we start our MC simulations from $12$ different independent initial conditions. We generate the $12$ initial conditions as follows. First, we keep all the monomers in two different arrangements inside a cylinder of length $108 a$ and diameter $14 a$ as shown in the Fig. \ref{fig111}. The two neighboring monomers along the chain contour remain at a distance $a$ from each other in the two arrangements. The spring interaction between the CL monomers is kept switched off. We keep the position of the first monomer of the chain fixed at one end of the cylinder since in bacteria {\em C. crescentus} the ori (origin of replication) is tethered at the boundary of the cylinder by suitable proteins, as is known from the experiments \cite{le}. In the generated initial configurations, the last monomer is not placed adjacent to the first monomer in both the cases, though for a ring  polymer the first and last monomers should be at a distance of $a$. Next, we switch on the only repulsive part of  the Lennard Jones potential (with a cutoff at $2^{1/6} \sigma$ and suitably shifted) between all the monomers, except the nearest neighbors along the chain contour. We evolve the system using Monte-Carlo simulations for $10^6$ MCS. To equilibrate the system we use following strategy: Initially we take a small value of spring constant between the first and the last monomer, i.e., $0.2 k_BT/a^2$ but keep the value of spring constant of the other springs (connecting the nearest neighbors along the contour) to be fixed at $200 k_BT/a^2$. We increase the value of the spring constant between the first and last monomer by $0.2 k_BT/a^2$ after every $1000$ MCS. After $10^6$ MCS, the spring constant will have the value $200 k_BT/a^2$ similar to the spring constant for nearest neighbors along the contour of the chain. 

To generate 12 different initial conditions, we do the following. Starting from the configuration at the end of $10^6$ MCS, we again equilibrate the system for a further $5\times10^6$ MCS with $6$ different random number seeds for each of the two configurations. After $5\times10^6$ MCS we get $12$ independent configurations of the ring polymer inside the cylinder, which we use as the initial conditions for the next set of runs from which we calculate statistically averaged quantities to  determine the organization of the polymer. We use these $12$ initial configurations as the starting points for our studies of a confined ring polymer for 3 cases: (a) ring polymer without CLs (b) ring polymer with CLs (c) ring polymer with CLs and weak attraction. Also, the first monomer (ori) is tethered at the boundary of the cylinder throughout the simulations and in all the independent runs for the model chromosome of bacteria {\em C. crescentus}. 

To ensure that the configurations we obtain after $5\times10^6$ MCS, are independent of each other we calculate the longest relaxation time of the polymer \cite{axel} by calculating the autocorrelation function of the  z-component of the vectors connecting monomer pairs numbered $1$ and $2008$, $1004$ and $3012$, $2510$ and $502$, $1506$ and $2514$ in $4017$ monomers chain in the presence of cylindrical confinement and with $\sigma=0.2a$. The axis of the cylinder lies along the z-axis. The autocorrelation function is calculated using the formula: 

$C_{ij}(w-w_{o})=\dfrac{\langle(z_{ij}(w)-\langle z_{ij}\rangle)(z_{ij}(w_{o})-\langle z_{ij}\rangle)\rangle}{\langle (z_{ij}(w)-\langle z_{ij}\rangle)\rangle^2}$

\begin{figure}[!hbt]
\includegraphics[width=0.7\columnwidth]{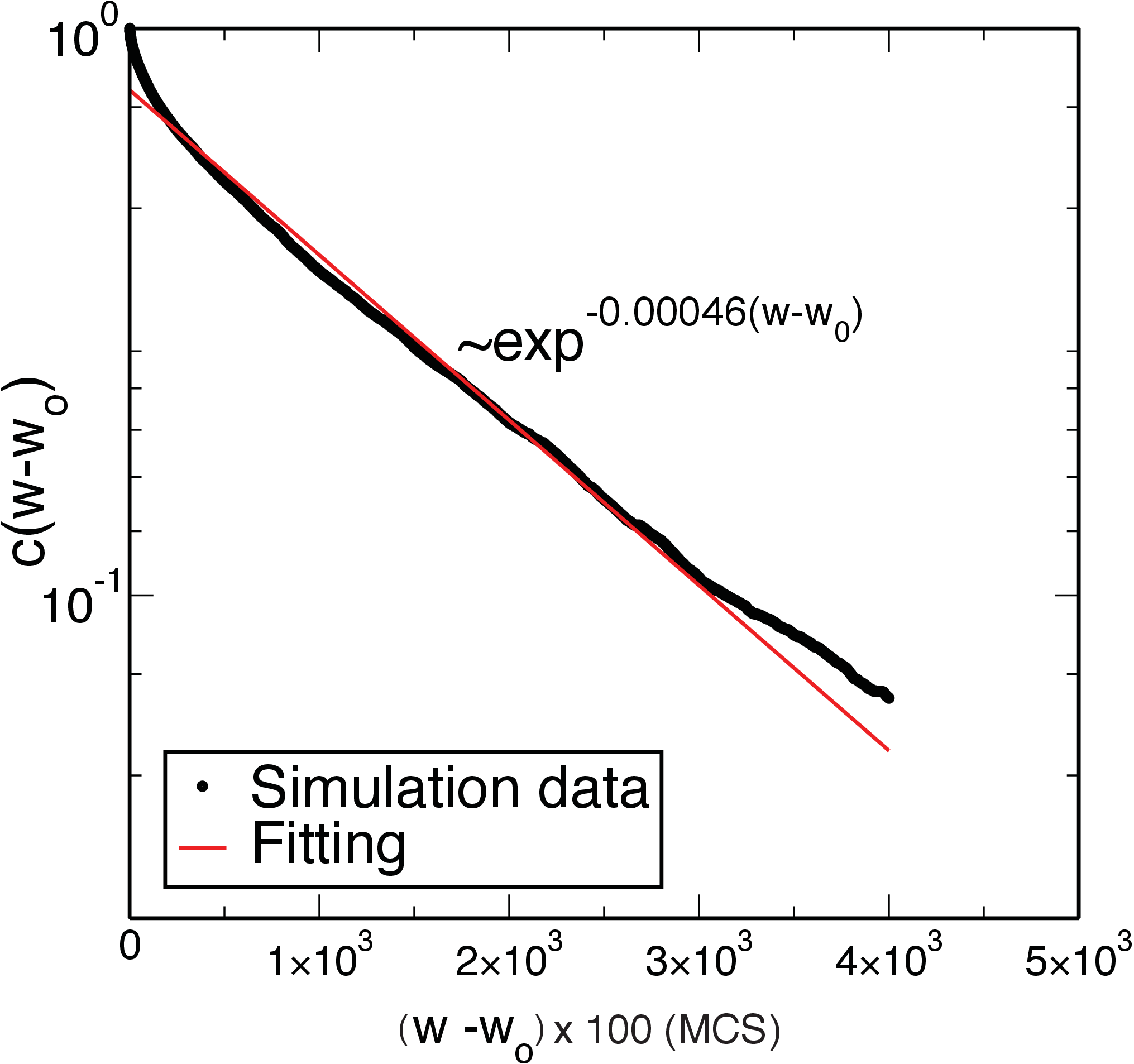}
\caption{\label{fig112}
The plot shows the average correlation function of the longitudinal component of the vector joining monomers numbered $1$ $\&$ $2008$, $1004$ $\&$ $3012$, $3012$ $\&$ $502$ and $1506$ $\&$ $2514$ for a polymer with $w-w_0$ in a semi-log plot. Each w represents $100$ MCS. The red line corresponds to the exponential fitting to the correlation function from the simulation. 
}
\end{figure}

Where $z_{ij}$ is the longitudinal component of the vector joining $i^{th}$ and $j^{th}$ monomer. The average $\langle .. \rangle$ is taken over MCS.  
Then we calculate the average value of the autocorrelation function of all the $4$ vectors. The function $C_{ij}(w-w_{o})$ is plotted in the figure Fig. \ref{fig112} versus $(w-w_o)$. Each $\Delta w=1$ corresponds to $100$ MCS, we collect data after every 100 MCS. For a ring polymer, we cannot calculate the standard end to end vector. Thus, we chose the vectors between the monomers which are at a maximum distance from each other along the chain contour.  We see from the graph that the correlation function of the longitudinal component of vector decays exponentially, and around $w-w_o=2000$ (i.e. $2\times 10^5$ MCS) the value of $C_{ij}(w-w_{o})$ is $1/e$ of its initial value.  Hence, we can expect the configurations we get after $5\times10^6$ MCS are independent of each other. 

\begin{figure}[!hbt]
\includegraphics[width=0.48\columnwidth]{caul_no_density_radial.eps}
\includegraphics[width=0.48\columnwidth]{caul_no_density_long.eps}
\includegraphics[width=\columnwidth]{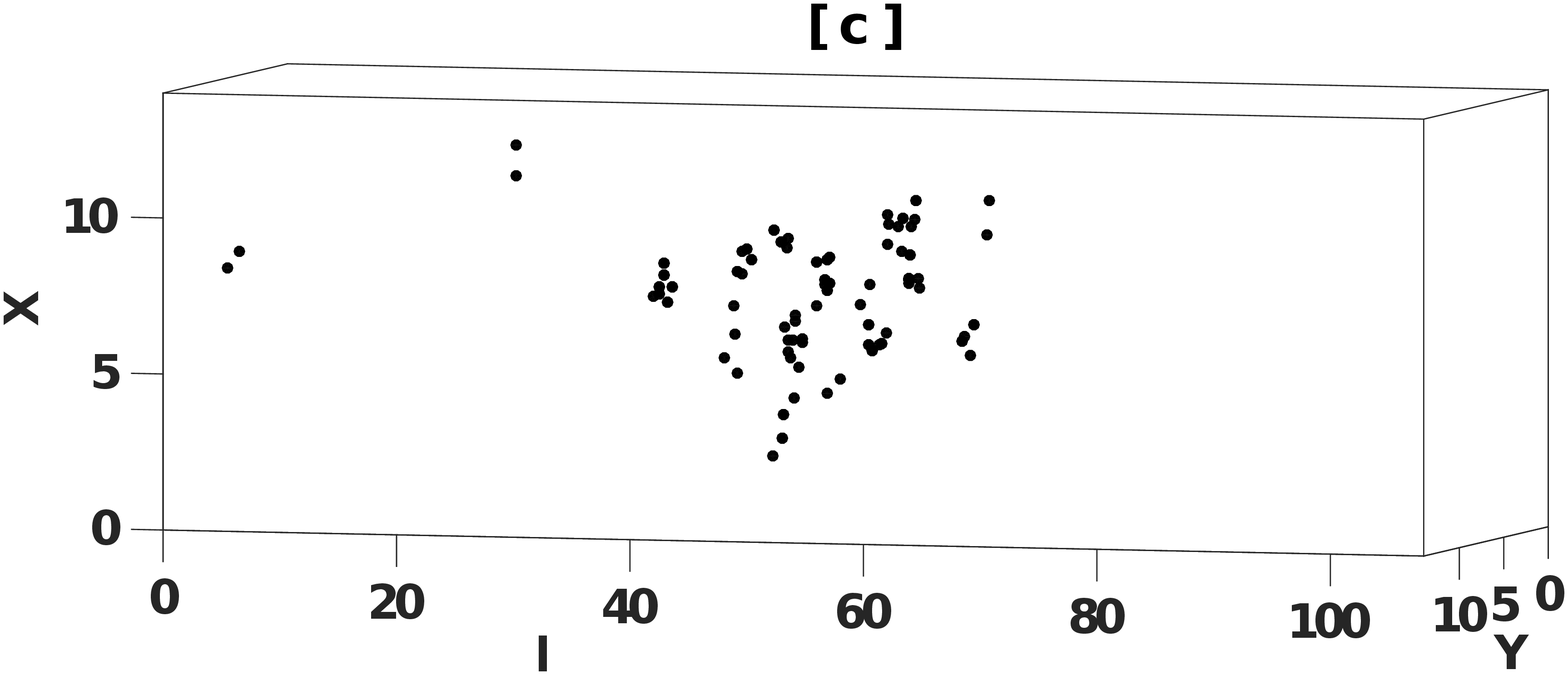}
\caption{\label{fig113}
The figures (a) and (b) show the number density of the monomers in the radial and longitudinal directions with the distance. Different lines correspond to the polymers with the different number of CLs. Error bars show the s.d. from the average value
across $12$ independent initial conditions. To get the radial number density, we calculate the number of monomers present between the concentric cylindrical shell ($dr=2a$) from the axis of the cylinder and divide it by the volume of the shell. The longitudinal number density represents the number of monomers within sections of the cylinder with radii $7a$ and length $2a$ divided by the total volume of the cylinder section. (c) Snapshot from the simulation shows the position of the monomers which constitute the CLs. The bounding box is given to read the positions of the CLs easily and does not represent the confinement geometry of the polymer which is a cylinder. 
}
\end{figure}

Now using these initial conditions, we can calculate the relevant statistical quantities to identify the organization of the DNA-ring polymer without any CLs.  To obtain statistical averages,  we further evolve the system for $29\times 10^6$ MCS using Monte-Carlo simulations, but start calculating the data statistical quantities after $5 \times 10^6$ MCS. Thus the data presented in Fig.\ref{fig113} and other  data for ring polymers without CLs is collected over $24 \times 10^6$ MCS for each initial condition,  where the data for averaging is taken after every $2\times 10^5$ MCS (since the polymer relaxes in about $2\times10^5$ MCS without CLs). 

Starting from the above mentioned initial configurations, we also introduce CLs to study the polymer configurations with $49$ no. of CLs (BC-1 CL-set) and $60$ CLs ($BC^{'}$ CL set). In the $12$ independent initial conditions, the positions of the monomers which constitute CLs can be greater than the bond-length $a$. Thus, we use the same strategy as mentioned previously to relax  the system.  Firstly, we take a small value of spring constant $\kappa_c=0.2 k_BT/a^2$ between the CL monomers and increase the value by $0.2 k_BT$ in every $1000$ MCS till it reaches the $\kappa_c=200 k_BT/a^2$.  This will lead to the cross-linked monomers to come near each other at the end of the equilibration, without affecting the stability of the computation.
We then evolve the system for a further $5\times10^6$ MCS for equilibration and then start  collecting data over the next $2.4 \times10^7$ MCS to calculate the average statistical quantities.  After that, we can compare the statistical quantities amongst independent runs (with different random number seeds for MC runs) starting from different initial conditions, and for the polymer with the different number of CLs. 

Firstly, we estimate the spread of the polymer inside the cylindrical confinement with and without CLs. For this, we first calculate the moment of inertia tensor of the polymer chain from its center of mass. Then we diagonalize the tensor to get the principal moments $I_1, I_2, I_3$, where $I_1 > I_2 > I_3$. The ratio of the principal  moments $I_1$ and $I_3$ gives the idea about the asymmetry of the polymer chain. Inside a cylinder, the chain is extended along the length of the cylinder and will be squeezed in the other directions. Hence the value of the ratio $I_1/I_3$ should be significantly greater than $1$. If the ring polymer is tethered at one end without any CLs,  the polymer should extend along the length  of the cylinder to maximize the entropy and  the value of the ratio $I_1/I_3$ is expected to be very large. But in the presence of the intra-chain cross-links, the polymer
should have less extension along the length because more constraints (CLs) lead to the compaction of the polymer. Thus, to get the estimate of the asymmetry and compaction of ring polymer along the direction of the cylinder's length we calculate the ratio $I_1/I_3$ for the three cases a) Ring polymer with no cross-links. b) polymer with $47$ CLs and c) polymer with $60$ CLs. The average values are given in the table \ref{table:1} where the average has been taken over the value obtained from $12$ independent initial conditions of the polymer. From the table, we see that with the increase in the number of CLs the value of $I_1/I_3$ decreases leading to the compaction of the polymer along the longitudinal direction of the cylinder, as expected. 
We also calculate the asphericity, $A_s=0.5\left[\dfrac{3(I_1^2+I_2^2+I_3^2)}{(I_1+I_2+I_3)^2}-1\right]$. The average values of $A_s$ for three cases are given in table \ref{table:1}. For a perfect sphere(rod) the value of $A_s=0(1)$. From the table, we see that as we increase the no. of CLs the value of $A_s$ decreases. This is because of the CLs compacts the polymer extension in the longitudinal direction. But the values of $A_s$ and $I_1/I_3$ does not change as we
increase the no. of CLs from BC-1 to $BC^{'}$.

\begin{table}[h!]
\centering
\begin{tabular}{ |p{1cm}| p{2.3cm}| p{1cm}|p{1cm}| }
\hline
No. of CLs  &  No. of {\em effective} CLs &  $I_1/I_3$  &   $A_s$ \\
\hline
0       &  0   &   50.38    &     0.23  \\
49      &  26   &  14.50     &    0.20   \\
60      &  33   &  13.89    &     0.19   \\
\hline
\end{tabular}
\caption{The table shows the average value of the moment of inertia ratio $I_1/I_3$ and the asphericity for a polymer with the different number of CLs. The average is taken over the configurations of polymer starting from $12$ independent initial conditions.}
\label{table:1}
\end{table}

Further to understand the distribution of the monomers inside the cylinder we calculate the number density of monomers in radial and longitudinal directions of the cylinder for the polymer with different number of CLs.  The average values of densities are  shown in the Fig. \ref{fig113}(a) and (b), where the average is taken over the value obtained from the $12$ independent initial conditions and the error bars show the s.d. from the average value. Different lines in the plot represent the number density of the polymer with the different number of CLs. From the plot, we see that the number density of monomers in the radial direction does not change with the increase in the 
number of CLs which suggests that the increase in the number of CLs is not helping in the compaction of the polymer in the radial direction. But we see that the number density of monomers significantly change in the longitudinal direction as we change the number of CLs. In the absence of any CLs, the polymer extends with an approximately uniform density of monomers  along the length of  the cylinder. But as we increase the number 
of CLs to $49$ the density at the center increases and gives a peak at the center of the cylinder. Further increasing the number of CLs from $49$ to $60$ there is a slight increase in the height of the peak. This suggests that the increase in the number of CLs help in the compaction of the polymer along the longitudinal direction while the radial directions remain unaffected.

We can also see that the longitudinal number density from $l=0$ to $l=26$ is very low, and the error bars are relatively smaller for the polymer with CLs. To investigate why this is the case, we plot the only the positions of the CL monomers from the simulation, which is shown in the Fig. \ref{fig113}(c). From the snapshot we see that the most of the CLs are clustered around $l=50$, thus giving a very high value of density at $l=50$. The number of monomers which constitute the CLs are relatively less in the region $l=0-26$ thus giving the value of number density and error bars to be less in that region.

\begin{figure}[!hbt]
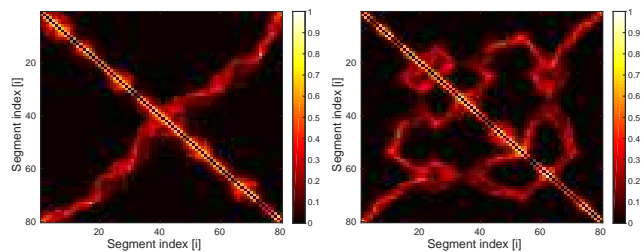

\includegraphics[width=0.48\columnwidth]{caul_init1_no_cl_exvol02_eps0.eps}
\includegraphics[width=0.48\columnwidth]{caul_init2_no_cl_exvol02_eps0.eps}
\caption{\label{fig4}
The colormaps show the probability of the CMs of different segments of the polymer to be within a distance $R_c<5a$ in the absence of any cross-links. 
The two colormaps are for the polymer starting with two different initial conditions, and the probability is an average, calculated over 
$120$ independent snapshots over $2.4 \times 10^7$ MCS for each run. }
\end{figure}

With this information, now we want to understand how does the internal organization of the polymer change in the presence of cylindrical confinement  with zero CLs, as well as with BC-1 and $BC^{'}$ set of CLs. For this, we calculate the positional correlation of different segments of the polymer with the different number of CLs. We can estimate the positional correlation as the probability of the CMs of the two segments ($50$ monomer in each segment) to be within cutoff $R_c$. A ring polymer with $4017$ monomers will have $80$ segments. We chose the value of cutoff $R_c$ to be $5a$ which we estimate according to the value of polymer's $R_g$ in the absence of confinement \cite{paper3}. The positional correlations are shown in Fig. \ref{fig4} as two representative colormaps for two independent runs. In the colormap, the x- and y-axis represent the segment index, and the color signifies the calculated probability. Bright color denotes the higher probability, and dark color corresponds to the lower probability. 

To explain the data of positional correlation of the colormaps for polymer without CLs, we define two parts of the ring polymer as arm-1 and arm-2. The arm1 consists of monomers numbered $1$ to $2008$, and arm2 consists of monomer numbered $2009$ to $4017$. The colormap of Fig. \ref{fig4} shows the positional correlation of the ring polymer under cylindrical confinement without any CLs. In the colormap of Fig. \ref{fig4} we see the two diagonals. The main diagonal (top-left of the figure to bottom-right) signifies that the segments which are neighbors along the contour of the chain are also spatially close to each other in the 3D space. For, e.g., segment number $40$ is close to segment number $38, 39, 41, 42$, etc. These correspond to the intra-arm-interactions of the segments of arm-1. The presence of other diagonal means that the monomers of arm-1 and arm-2 are coming closer to each other. This suggests that the two arms  are arranged parallel to each other along the long axis of the cylinder.  These diagonals are also present in the experimental contact map of the bacteria {\em C. crescentus} and suggest that the two arms of the chromosome are arranged parallel to each other in-vivo as was also reported in previous study \cite{le}. But by comparing the simulation colormap with the experimental colormap, we observe that the off-diagonal points which are present in the experimental contact map, are missing in the simulation colormap. Also, the colormaps from $12$ independent conditions do not match with each other as can be seen from the two representative colormaps of Fig. \ref{fig4}. This suggests that the confinement of the cell helps in arranging the two arms of the polymer to arrange parallel to each other but does not organize the polymer completely into a particular structure as different initial conditions show the different positional correlation of the segments.

\begin{figure}[!hbt]
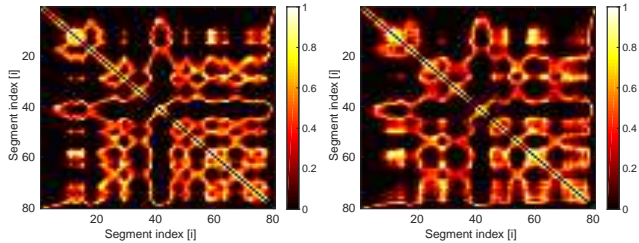

\includegraphics[width=0.48\columnwidth]{caul_init1_BC1_exvol02_eps0.eps}
\includegraphics[width=0.48\columnwidth]{caul_init2_BC1_exvol02_eps0.eps}
\caption{\label{fig5}
The colormaps show the probability $p(i,j)$ of the CMs of two different segments $i,j$ of the
polymer to be within a distance $R_c<5a$ with BC-1 set of cross-links ($49$ CLs). The two colormaps are for the polymer starting with two different initial conditions.
}
\end{figure}

\begin{figure}[!hbt]
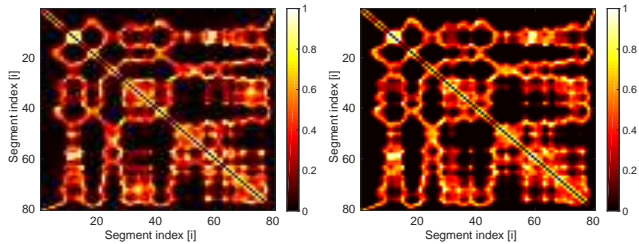

\includegraphics[width=0.48\columnwidth]{caul_init1_60_CLs_exvol02_eps0.eps}
\includegraphics[width=0.48\columnwidth]{caul_init2_60_CLs_exvol02_eps0.eps}
\caption{\label{fig6}
The colormaps show the probability $p(i,j)$ of the CMs of two different segments $i,j$ of the
polymer to be within a distance $R_c<5a$ with $BC^{'}$ set of cross-links ($60$ CLs). The two colormaps are calculated from two independent MC runs, starting with two different initial configurations of the polymer.}
\end{figure}

We next investigate how is cross-linking the specific pairs of monomers help in the organization of the polymer? Will be able to obtain off-diagonal bright pixels in the positional correlation colormaps? And  will the positions of the bright pixels obtained from  simulations match with those observed  in the (coarse-grained) experimental contact maps? We reiterate that we cross-link $49$  (BC-1 CL set) and $60$ ($BC^{'}$ CL-set) pair of specific monomers, which are $26$ and $33$ effective CLs, respectively. Note, that the number of cross-links is significantly less than the number of CLs in BC-2 CL-set ($153$ CLs)  which we used in our previous studies \cite{epl,jpcm} to obtain the organization of DNA-polymer without confinement. The two representative colormaps using data from two independent MC runs starting from different initial conditions are shown in Fig. \ref{fig5} and Fig. \ref{fig6}  for BC-1 and $BC^{'}$ CL sets, respectively.

From the figures, we see that in the colormap of the polymer with BC-1 CL set we get  the off-diagonals pixels as well as the bright diagonal  pixels, which are also present in the experimental contact map \cite{epl}. We also quantify the number of bright pixels in the diagonals, data for which is given later in this section. Also, the colormaps from  the $2$ independent initial conditions look statistically similar. We also check the colormaps from $12$ initial conditions and all the colormaps look statistically similar as the two in Fig. \ref{fig5}. But on increasing the number of CLs from $49$ to $60$, we see that colormap does not change significantly, though we, of course get additional bright patches in the colormap. Hence, we can claim that $60$ CLs ($33$ effective CLs) can organize the polymer into a particular structure. Also, note that in the absence of confinement the colormaps were predominantly black with very less number of bright pixels present in the colormap for BC-1 CL set \cite{jpcm,epl}. But in the presence of confinement only $60$ CLs are enough to organize the polymer into a unique structure. We check the colormaps from all the $12$ independent initial conditions and they all look statistically similar for the polymer with both BC-1 and $BC^{'}$ CL set,  relevant Pearson correlations quantifying the above claim is presented at the end of this section.

Now we present our result for the polymer with $BC^{'}$ CL set, cylindrical confinement, but now also incorporating a small attraction between the monomers using the LJ potential mentioned in the model section with the attraction strength parameter $\epsilon=0.3 k_BT$. Starting from the $12$ initial conditions for  ring polymers without CLs, we introduce the weak attraction at the same time as we introduce the CLs. Before starting to collect data to calculate the average of statistical quantities, we  allow the polymer to relax using the same protocol as described before for the polymers with CLs, but without the weak attraction.  

\begin{figure}[!hbt]
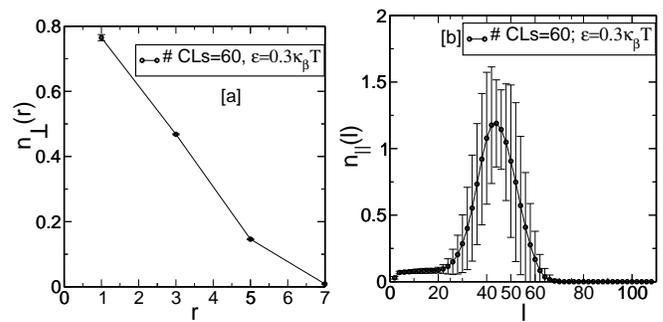

\includegraphics[width=0.49\columnwidth]{caul_no_density_radial_attarction.eps}
\includegraphics[width=0.49\columnwidth]{caul_no_density_long_attraction.eps}
\caption{\label{fig10}
The plots (a) and (b) show the number density of the monomers in the radial and longitudinal directions with the distance in the presence of attraction between the monomers. Error bars show the s.d. from the average value across $12$ independent initial conditions. 
}
\end{figure}

As mentioned in the introduction, the small attraction between the monomers incorporates the effect of molecular crowding. The attraction between the monomers should further help in the compaction of the polymer chain. To check whether this is the case, first we calculate the number density of the monomers in radial and longitudinal directions and then we calculate the values of the ratio of principal moment of inertia, $I_1/I_3$ and asphericity, $A_s$ and compare it with the polymer with $BC{'}$ CLs under confinement but no attraction between the monomers.
The no. densities in the radial and longitudinal directions are shown in the Fig. \ref{fig10}. On comparing the no. densities in Figs.\ref{fig113} and \ref{fig10} (a) we see that the attraction between the monomers helps in the compaction of the polymer in the radial directions as well. Note that, on increasing the number of CLs the radial density was not changing. But the small attraction between the monomers helps the polymer to get compact in the radial as well as in the longitudinal directions.

Further, in the presence of attraction the average values of the ratio $I_1/I_3$ and $A_s$  are $11.38$ and $0.18$ respectively. On comparing with the values in table \ref{table:1} for the polymer with $BC^{'}$ set of CLs, we see that the attraction between the monomers indeed compact the polymer and decreases its asphericity. 

\begin{figure}[!hbt]
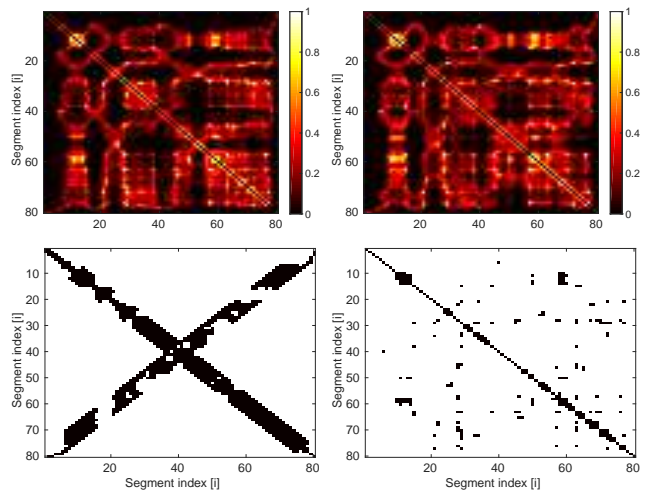

\includegraphics[width=0.48\columnwidth]{caul_init1_60_CLs_exvol02_eps03.eps}
\includegraphics[width=0.48\columnwidth]{caul_init2_60_CLs_exvol02_eps03.eps}
\includegraphics[width=0.48\columnwidth]{diagonalp005.eps}
\includegraphics[width=0.48\columnwidth]{offdiag05.eps}
\caption{\label{fig7}
The top two figures show the positional correlation colormaps for the center of mass of different polymer segments in the presence of small attraction between the monomers and with $BC^{'}$ set of CLs. The value of the cutoff $R_c$ to obtain the colormap is set as $R_c = 4a$ The two colormaps are from the runs starting from two different initial conditions. The bottom two figures: for aid visualization of the diagonal and off-diagonal pixels we plot only the diagonal pixels with $p>0.05$ of the positional correlation maps in the bottom left plot and off-diagonal pixels with $p>0.5$ in the bottom right plot in the binary form in a binary black and white format. These can now be easily compared with the coarse grained experimental contact map (refer supplementary section). The data to generate the coarse grained experimental contact maps have been downloaded from \cite{le}}.
\end{figure}

To investigate the internal organization of the polymer in the presence of small attraction between monomers with the value of parameter $\epsilon=0.3k_BT$, we calculate the positional correlation of different segments as before. To estimate the value of $R_c$ we take the help of our study of the polymer with different attraction strength parameter $\epsilon$, with BC-2 sets of CLs and without confinement. The value was taken to be $R_c=4a$. The positional correlations are represented as a colormap in Fig. \ref{fig7}(Top). The two colormaps correspond to the polymer starting from the $2$ independent initial conditions.
From the colormaps, we see that the two colormaps look statistically similar. Also, the overall positional correlation in the colormaps of Fig. \ref{fig7} do not look significantly different from the colormaps of the Fig. \ref{fig6}. But note that the cutoff $R_c$ we chose to calculate the positional correlation has been decreased to $4a$ in this case, which was chosen to be $R_c=5a$ to plot the colormap of Fig. \ref{fig6}. So, the small attraction between the monomers leads to overall compaction of the polymer but does not change the internal organization of the polymer significantly in the presence of cylindrical confinement. Also, for the better visualization of the off-diagonal and diagonal pixels in the top two simulation colormaps of Fig. \ref{fig7} we plot the diagonal pixels with $p>0.05$ and off-diagonal pixels with $p>0.5$ separately in the bottom two figures of Fig. \ref{fig7} in the binary form. On comparing the bottom two figures with the experimental coarse-grained colormap of Fig. S1 we see that the diagonal and off-diagonal pixels which are present in the coarse-grained experimental contact map are also present in the simulation colormap. We also quantify the number of pixels along the Diagonal-2 later in this section and compare it with the positional correlation colormaps of polymer with the BC-2 set of CLs and no confinement. 

Next, we check for the statistical equivalence of organization obtained from $12$ independent MC simulations as observed  in the colormaps. Thus, we calculate the average value of the Pearson correlation of the positional correlation colormaps for the polymer (a) with no CLs (b) with BC-1 set of CLs (c) with $BC^{'}$ set of CLs and (d) with $BC^{'}$ set in the presence of small attraction between the monomers . Corresponding to runs starting from the $12$ initial conditions there will be ${}^{12}C_2=66$ values of Pearson correlation (pc) for each pair of graphs. We calculate the average value of Pearson correlation $\langle pc \rangle$ for each of the $4$ different studies. The values of $\langle pc \rangle$ have been listed in the table \ref{table:2}.  For details of the calculation procedure of  $\langle pc \rangle$, refer \cite{paper3}.

\begin{table}[h!]
\centering
\begin{tabular}{ |p{1.2cm}| p{2cm}|p{1cm}| }
\hline
No. of CLs  & Monomer Attraction  & $<pc>$ \\
\hline
0       &  absent   &   0.51  \\
49      &  absent  &    0.54   \\
60      &  absent   &   0.60  \\
60      &  present  &   0.70   \\
\hline
\end{tabular}
\caption{The table shows the average value of Pearson correlation $<pc>$ for polymer with different value of CLs and in the presence and absence of attraction between the monomers. Note, to calculate the value of pc for a pair of colormaps, we consider only those pixels corresponding to segments $i$ and $j$, for the probability $p(i,j)>0.05$ in at least one of the colormaps from different runs. This is to avoid the bias in calculating pc, from the large number of pixels which are completely dark.}
\label{table:2}
\end{table}

From the table \ref{table:2}, we see that the value of Pearson correlation increases as we increase the number of CLs. We get the relatively higher value of $\langle pc \rangle$ if we introduce the small attraction between the monomers.

Previously, we have mentioned that cylindrical confinement helps in the parallel arrangement of two arms of the polymer and thus we get another diagonal (Diagonal-2) in addition to the main diagonal (Diagonal-1) in the experimental contact-map of bacteria {\em C. crescentus}. In our previous study of the DNA ring-polymer with BC-2 set of CLs ($159$ number of CLs) and in the absence of the confinement we did not obtain the bright pixels along the Diagonal-2 as seen in the coarse-grained experimental contact map of {\em C. crescentus}(See supplementary material FigS1) though there was a good match of the off-diagonal points in the experimental and simulation contact maps \cite{epl}. But in the current study, in the presence of confinement, we see that the bright pixels are also present along the Diagonal-2 in the colormaps of Fig. \ref{fig7}. To quantify the presence of pixels along the Diagonal-2, we calculate the number of bright pixels ($n_d$) with $p>0.05$, normalized suitably along the Diagonal-2 which are present in the colormap of Fig. \ref{fig7} and compare it with the positional correlation colormap of polymer with the BC-2 set of CLs with no confinement in \cite{epl}. We choose the pixels along the diagonal-2 such that it consists of $5$ nearest neighboring pixels (in the vertical and horizontal direction) of a particular segment index in the simulation colormaps, similar to the coarse-grained experimental contact map in the Fig. S1 of the supplementary section. We find that the average value of $n_d$ is significantly high, i.e., $n_d=0.68$  for the colormaps of Fig.  \ref{fig7} compared to the relatively smaller value of $n_d$ equal to $0.44$ in the colormap of Fig. 2(a) in the \cite{epl}.  

\begin{figure}[!hbt]
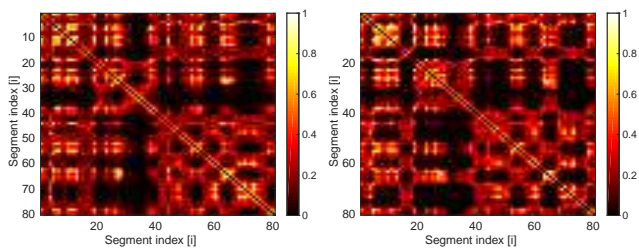

\includegraphics[width=0.48\columnwidth]{ecoli_eps03_77CLs_exvol02_init1.eps}
\includegraphics[width=0.48\columnwidth]{ecoli_eps03_77CLs_exvol02_init4.eps}
\caption{\label{fig100}
The figure show the positional correlation colormaps for the center of mass of different polymer segments in the presence of small attraction between the monomers and with $77$ number of CLs taken from the contact map of bacteria {\em E. coli}. The value of the cutoff $R_c$ to obtain the colormap is set as $R_c = 4a$. The two colormaps are from the runs starting from two different initial conditions. }
\end{figure}

In the figure \ref{fig100}, we also plot the positional correlation colormaps corresponding to the model chromosome of bacteria {\em E. coli} in the presence of small attraction between the monomers with strength $\epsilon=0.3k_BT$ using the same procedure that was used to calculate the colormaps of {\em C. crescentus}. 
The two colormaps are in Fig. \ref{fig100} are from the independent MC runs starting from the two different initial conditions. We again see that the colormaps from independent runs look statistically similar to each other and the average value of $\langle pc \rangle=0.75$ is relatively high across independent runs starting from $12$ initial conditions. Also, on comparing the simulation colormaps of \ref{fig100} with the experimental coarse-grained colormap of Fig. S1 in the supplementary section, we see that all the pixels which are present in the experimental contact map are also present in the colormaps obtained from our simulations. But note, now we have used fewer CLs (i.e., $77$ CLs, effectively $38$ CLs) compared to the BC-2 set of CLs (i.e., $159$ CLs, effectively $82$ CLs).

\begin{figure*}[t!]
\includegraphics[width=1.6\columnwidth]{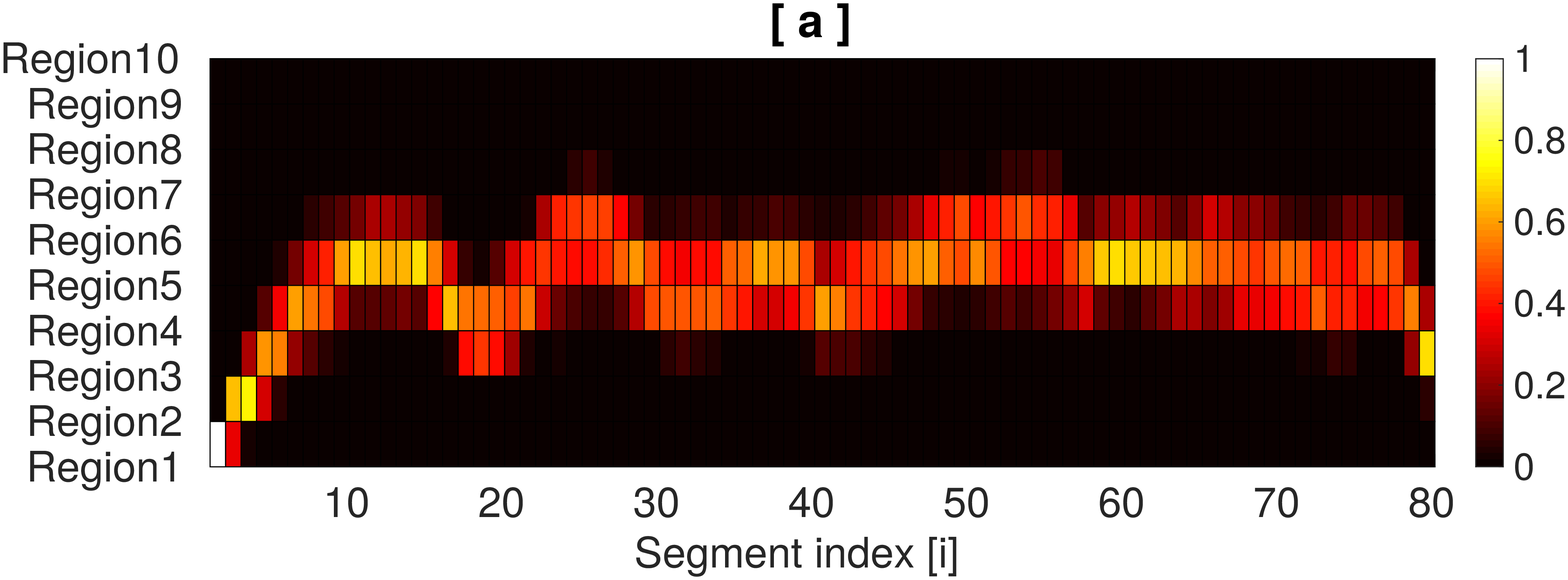} \\
\hskip1cm
\includegraphics[width=1.6\columnwidth]{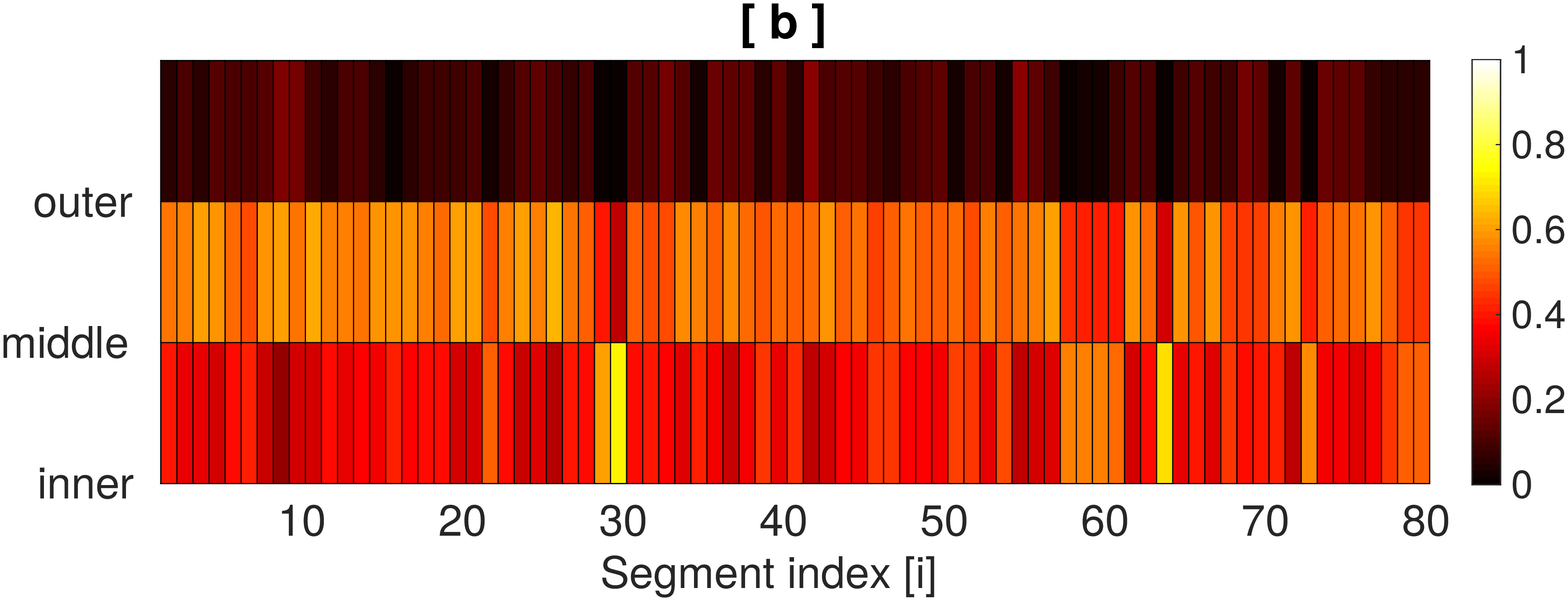} \\
\hskip1cm
\includegraphics[width=1.6\columnwidth]{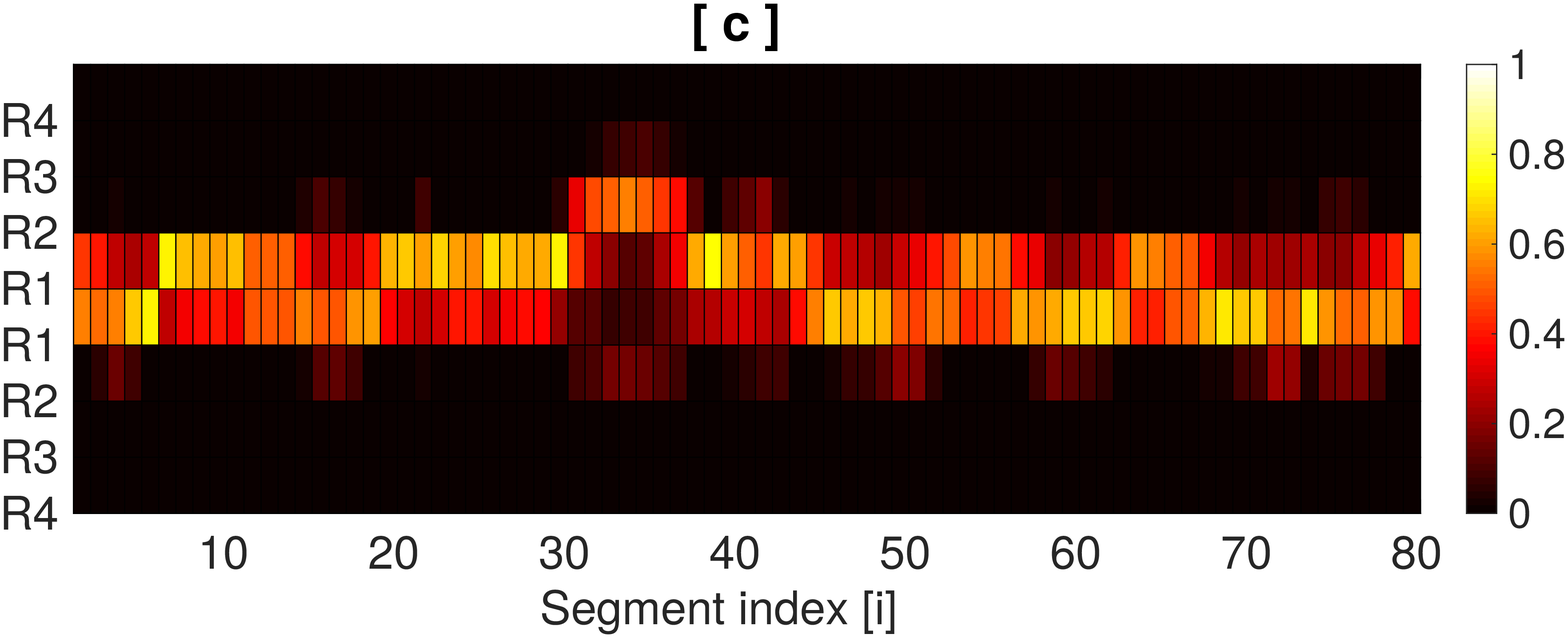} \\
\hskip1cm
\includegraphics[width=1.6\columnwidth]{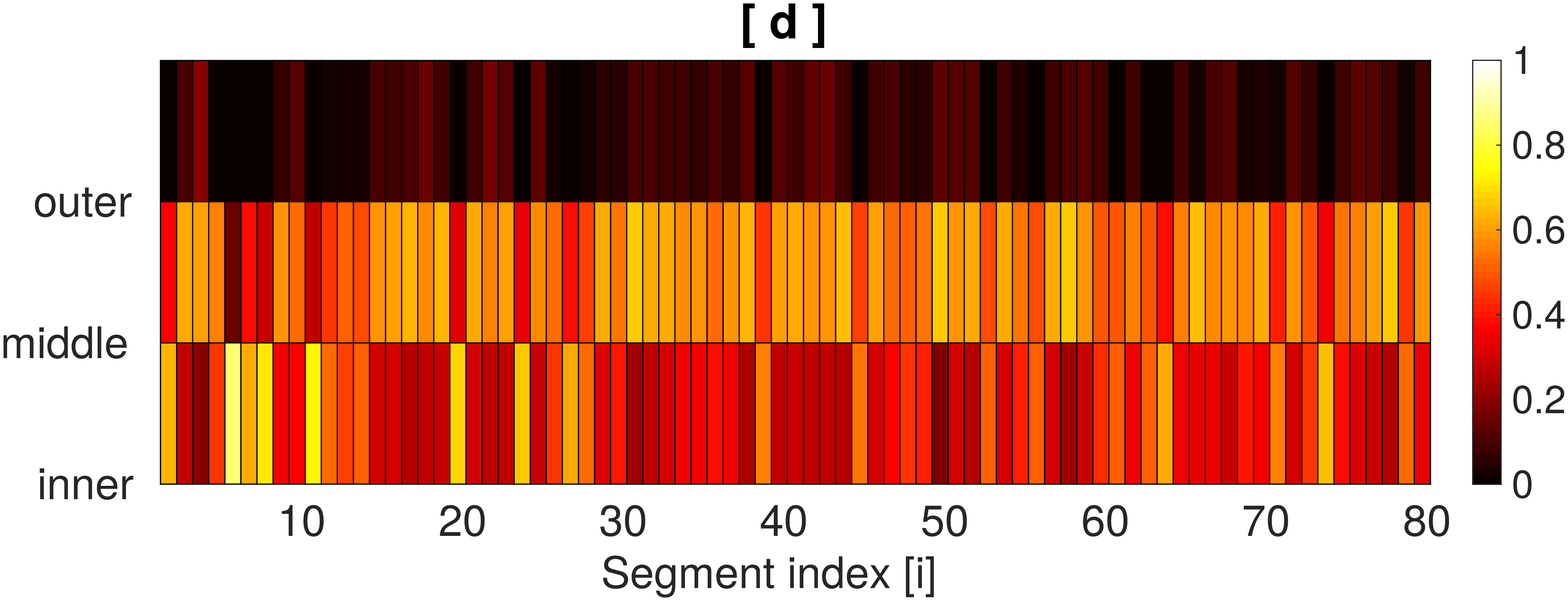}
\caption{\label{fig8}
The colormaps (a),(c) show the longitudinal  and (b), (d)  the radial distributions of the positions of the center of mass of different segments for model chromosome of bacteria {\em C. crescentus} and {\em E. coli}, respectively. The color bar denotes  the probability to find a segment in a particular region. Refer text for description of how the different regions are specified. In essence, this figure gives our prediction of the 3D organization of the different segments of the DNA-polymer within the cylinder for both the bacteria {\em C. crescentus} and {\em E. coli} using the thesis that few CLs at specific positions along the chain contour, effective attraction between the monomers and confinement effects are enough to organize the bacterial chromosome in the space within the cell.
}
\end{figure*}

Now, to estimate the overall organization of the polymer in the presence of CLs at specific positions, confinement and the small attraction between the monomers we calculate the radial and longitudinal distribution of positions of the center of mass of  different segments for two bacterial chromosomes.
Both the longitudinal and radial distribution of positions of CMs
of different segments is averaged over the runs starting from $12$ independent initial condition. The longitudinal distribution of the positions of CMs of the segments is shown as the colormap in Fig. \ref{fig8} (a), (c) for {\em C. crescentus} and {\em E. coli} ,respectively.  The color shows the probability of the z-component of a particular segment to be found in the region numbered $1,2..,10$ for {\em C. crescentus} or Left-R1, Left-R2..   for {\em E. coli}.  For the model chromosome of bacteria {\em C. crescentus}, we measure the position from the location of monomer numbered $1$, which is tethered at one end of the cylinder. For the bacteria {\em E. coli} the ori is not tethered at the boundary of the cylinder. Thus we measure the longitudinal distribution of the segments from the center of mass of the longitudinal component (z-component). 

In Fig. \ref{fig8} (a) the x-axis represents the segment index, and the y-axis represents the different regions numbered $1,2,..5$ along the length of the cylinder starting from the ori position. Each region corresponds to a length of $l=10a$, i.e., Region1 corresponds to $l=0-10a$, Region2 corresponds to $l=20a-30a$ and so on. In Fig. \ref{fig8}(c) x-axis shows the segment index $i=1-80$ and the y-axis represents the different regions measured from the center of mass of the polymer. The region left-R1 correspond to the region at a value of z-component between  $0$ and $-10$, the region right-R1 represents the region at a value of z-component between $0$ and $+10$  from the center of mass.

For the radial distribution, we define three regions inner ($r\leq2a$), middle ($2a <r\leq4$)and outer ($r>4a$) from the the axis of the cylinder as  monomer number 1 is tethered at the axis of the cylinder for the bacteria {\em C. crescentus} at one end of the cylinder. For the bacteria {\em E. coli}, where the  ori is not tethered at the boundary of the cylinder, we define the three regions from the center of mass of radial components ($x$ and $y$). We then calculate the probability of radial component of the CM of the segments to be found in the inner, middle and outer region. Then we average the probability over $12$ independent runs starting from the $12$ different initial conditions. It is shown as the colormap in Fig. \ref{fig8}(b) and (d). The x-axis shows the segment index and the y-axis correspond to the three regions inner, middle and outer. The color shows the probability of a segment to be found in the three regions. 

The positional colormaps, the longitudinal and radial distribution of the polymer segments give the overall 3D organization of the model chromosome. In our previous work, we observed that for the model chromosome of bacteria {\em E. coli} the polymer segments with highly expressed genes were found in the outer region of the globule with high probability. 
\clearpage
Here also, we check that the chromosome segments of bacteria {\em E. coli} which consist of highly expressed genes are found in the middle and outer regions, and they have a very less or zero probability to be found in the inner region. Hence, our results are also consistent with our previous prediction of the organization of the chromosome without confinement.

\begin{figure}[!hbt]
\includegraphics[width=\columnwidth]{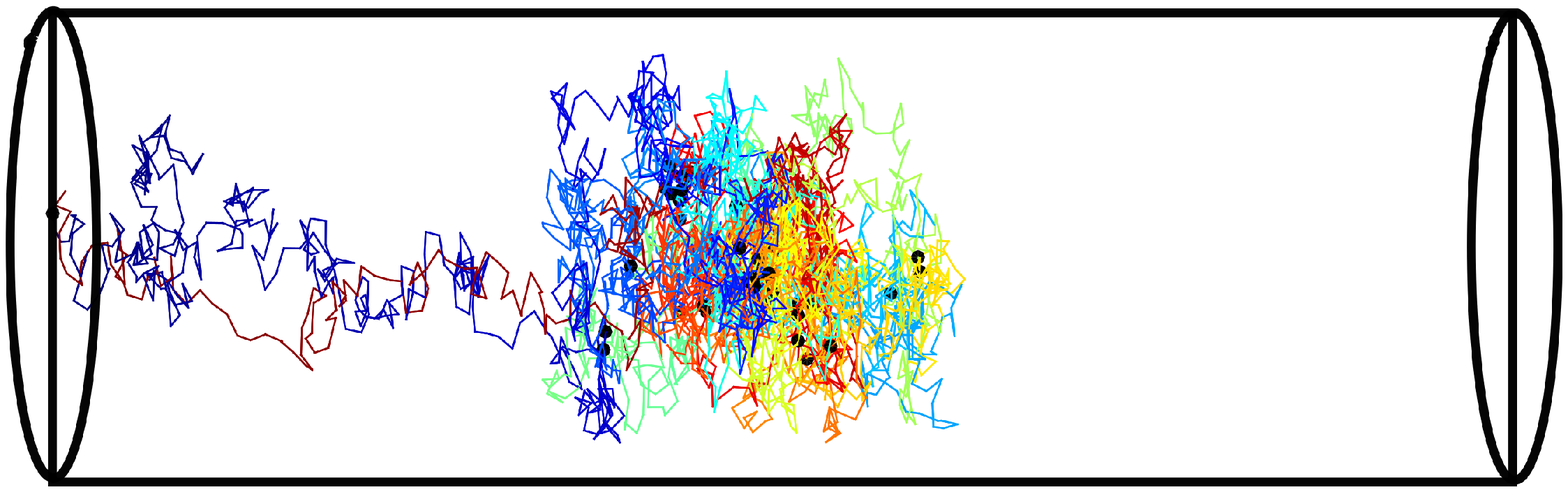} \\
\includegraphics[width=\columnwidth]{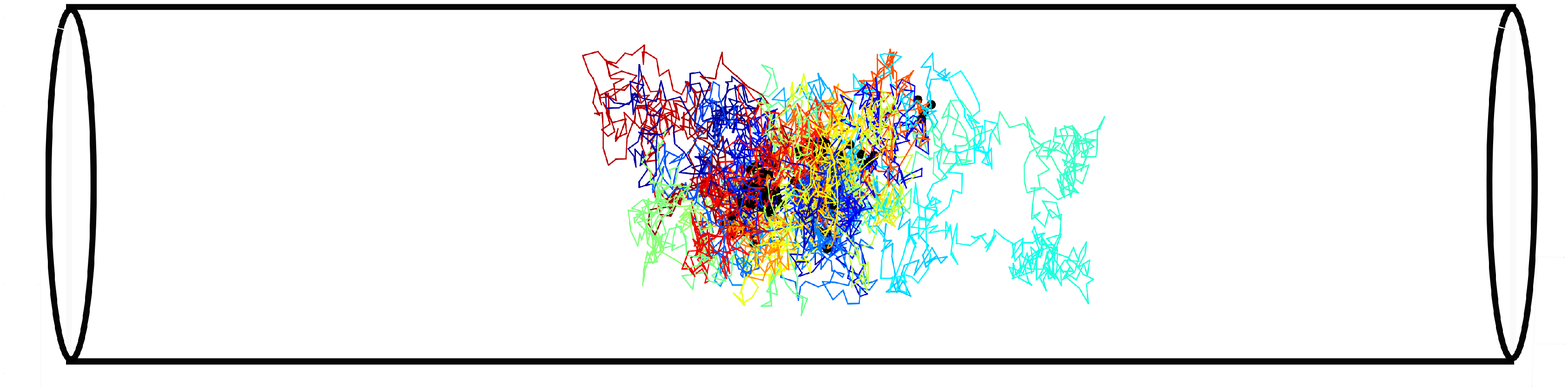} \\
\caption{\label{fig102}
The snapshots from the simulation for the model chromosome of bacteria {\em C. crescentus} and {\em E. coli}. The positions of CLs are represented by the black points.
}
\end{figure}

We also present the snapshots from the simulations in the Fig. \ref{fig102} for the model chromosomes of bacteria {\em C. crescentus} and {\em E. coli}. The monomers are colored from blue to red  according to their positions along the chain contour. For the model chromosome of {\em E. coli} where the first monomer is not tethered to the boundary of the cylinder, the 
polymer occupies the center of the cylinder whereas for the chromosome of {\em C. crescentus} we see that most of the monomers occupy the center of the cylinder while the first monomer is tethered at the boundary of the cylinder. We also mark the positions of the CL monomers with black color. From the snapshots, we can confirm that the CLs are clustered in space for both the model chromosomes.
In this study, we took a larger aspect ratio ($1:7.5$) of the cylinder compared to the aspect ratio ($\approx 1:5$ as used by other researchers for their studies \cite{le,william}. Even though we use a longer cylinder (of length $108a$ and rather than the expected $70a$ for a cylinder radius of $14a$), we observe that the polymer is collapsed in the cylinder, and the monomer density is zero beyond $75a$.  For {\em C. crescentus}, the ori is fixed at one end of the cylinder, whereas the ori for {\em E. coli} is at the center of the coil. The special position of the {\em C. crescentus} ori contributes to the presence of the diagonal seen in the contact map for {\em C. crescentus}. The consistency of the {\em E. coli} organization with our previous results means that the gene-dense regions and the active regions are found on the peripheral of the chromosome, as was pointed out earlier in \cite{epl}. This is expected to have important biological consequences.

In our studies, we have considered the cross-linked monomers to be at fixed positions along the contour of the polymer. We note that other studies \cite{nicodemi,geoffrey,marenduzzo} consider the binders (cross-links) between different segments of the chain to have the ability to diffuse around as is known to be the case experimentally. But we have considered only $33$ ($\& 38$) {\em effective} cross-links in a $4$ ($\& 4.6$) million base pair DNA chain, modeled with $4$ (and $4.6$) thousand beads in a bead-spring polymer model. Other than the ones we have considered, there can be other binding proteins which can diffuse around and play a role in the local and more detailed organization. Secondly, even if the position of the cross-links in our study moves by  ]few ($4$ or $5$) monomers on either side of its present position, the obtained organization cannot be significantly different, within the statistical fluctuations. As each monomer represents $1000$ base pairs, a variation of the position of the cross-link over $8$ monomers represents a diffusion of the DNA-binding protein along the contour over $\approx 8$  kilo base pairs.

\vskip 4cm

\section{Discussions}
We showed the organization of bacterial DNA-polymers  in cylindrical confinement and in the presence of optimally chosen weak attraction between the monomers with the different number of cross-links at specific positions along the chain contour. These specific cross-links are taken from the contact maps of bacterial DNA. In our previous studies, we had shown that in the absence of confinement and molecular crowders, we were able to get the structure of the DNA-polymer with $159 \& 153$ cross-links ($82 \& 60$ effective cross-links) \cite{epl,jpcm} for {\em E. coli} $\&$ {\em C. crescentus}, respectively. In this study, we show that we get the organization of the DNA polymer with only $38 \& 33$ effective cross-links, respectively, for both the bacterial chromosomes when we incorporate the effects of molecular crowders, ability to release topological constraints and confinement. So, in effect, we show that the very few specific cross-links, confinement and the effect of molecular crowders play the pivotal roles for the polymer to 
acquire a particular organization. The Pearson correlation on comparing positional-correlation contact-maps from independent MC runs (starting from the independent initial conditions) are $=0.7$ for {\em C. crescentus} and $0.75$
for {\em E. coli}. The ability to release topological constraints also plays a crucial role in the organization the DNA polymer.  

We obtain a better agreement of our predicted organization with the coarse-grained experimental contact map of 
{\em C. crescentus}, viz., the pixels along the secondary diagonal of the positional correlation colormaps  which were absent in our previous studies \cite{epl}  are now present for the polymer in the presence of $BC^{'}$ set of cross links, confinement and crowders. The organization of the chromosome of {\em E. coli} in the presence of these added effects  are consistent with our previous prediction of the organization of the chromosome, which in turn  was in agreement with the coarse-grained experimental contact map for {\em E. coli}. But, now we can predict the approximate positions of the center of mass of different polymer segments each consisting of $50 \& 58$ monomers ($50 \& 58$ kilo base pairs) for {\em C. crescentus} and {\em E. coli}, respectively,  in the 3D space within the cylinder. We hope that our prediction can be validated in the future experiments.

In our prediction of the 3D organization, we do not pin-point the exact position of a segment within the cylindrical cell. But, our predictions of the 3D organization of DNA polymer lists the segments which are likely to be found in the peripheral regions with higher probability or the segments which are likely to be found in the central core region. Moreover, for {\em C. crescentus}, we predict the relative distance from the ori which is fixed at one end of the cell. However, for {\em E. coli}, we measure the distance from the center 
of the coil (center of mass), and from the data, we can experimentally validate which segments are likely to be opposite each other in space along the axis. We further observe in our simulations that the CLs are clustered at the center of the cylinder and the coil, this is consistent with the prevalent understanding that DNA organization at large length scales in bacteria is seen by formation of long loops "emanating" out from  clusters of DNA-binding proteins \cite{amos}.

We were pleasantly surprised that relatively very few cross-links can organize a long ring polymer. Of course for DNA, at smaller length scales Nucleoid associated proteins (NAP) is expected to play a role in locally compacting the DNA-chain at the scale of $\approx 150$ base pairs or around $20-30$ nm. We believe our model much simpler than other present models which optimize multiple parameters to predict the contact map. The only parameter we have optimized in our the preceding manuscript and used here is the weak attraction $\epsilon=0.3 k_BT$, which accounts for the role of crowders and the ability of the chains to cross themselves.

\section{acknowledgments}
%We thank Farhat Habib and G.P. Manjunath for introducing this research problem to us and discussions during our collaboration which resulted in the papers \cite{jpcm,epl}. Their guidance and analysis of contact map data was crucial for our previous  investigations,  and without which we could not have completed this present work.
We acknowledge the use of computer cluster bought from DST-SERB Grant No.  EMR/2015/000018  and  funding  from  DBT  Grant BT/PR16542/BID/7/654/2016 to A. Chatterji. AC. acknowledges funding support by DST Nanomission, India under the Thematic Unit Program (Grant No. SR/NM/TP-13/2016). 
\bibliographystyle{apsrev4-1}
\bibliography{paper4}

\begin{thebibliography}{31}%
\makeatletter
\providecommand \@ifxundefined [1]{%
 \@ifx{#1\undefined}
}%
\providecommand \@ifnum [1]{%
 \ifnum #1\expandafter \@firstoftwo
 \else \expandafter \@secondoftwo
 \fi
}%
\providecommand \@ifx [1]{%
 \ifx #1\expandafter \@firstoftwo
 \else \expandafter \@secondoftwo
 \fi
}%
\providecommand \natexlab [1]{#1}%
\providecommand \enquote  [1]{``#1''}%
\providecommand \bibnamefont  [1]{#1}%
\providecommand \bibfnamefont [1]{#1}%
\providecommand \citenamefont [1]{#1}%
\providecommand \href@noop [0]{\@secondoftwo}%
\providecommand \href [0]{\begingroup \@sanitize@url \@href}%
\providecommand \@href[1]{\@@startlink{#1}\@@href}%
\providecommand \@@href[1]{\endgroup#1\@@endlink}%
\providecommand \@sanitize@url [0]{\catcode `\\12\catcode `\$12\catcode
  `\&12\catcode `\#12\catcode `\^12\catcode `\_12\catcode `\%12\relax}%
\providecommand \@@startlink[1]{}%
\providecommand \@@endlink[0]{}%
\providecommand \url  [0]{\begingroup\@sanitize@url \@url }%
\providecommand \@url [1]{\endgroup\@href {#1}{\urlprefix }}%
\providecommand \urlprefix  [0]{URL }%
\providecommand \Eprint [0]{\href }%
\providecommand \doibase [0]{http://dx.doi.org/}%
\providecommand \selectlanguage [0]{\@gobble}%
\providecommand \bibinfo  [0]{\@secondoftwo}%
\providecommand \bibfield  [0]{\@secondoftwo}%
\providecommand \translation [1]{[#1]}%
\providecommand \BibitemOpen [0]{}%
\providecommand \bibitemStop [0]{}%
\providecommand \bibitemNoStop [0]{.\EOS\space}%
\providecommand \EOS [0]{\spacefactor3000\relax}%
\providecommand \BibitemShut  [1]{\csname bibitem#1\endcsname}%
\let\auto@bib@innerbib\@empty
%</preamble>
\bibitem [{\citenamefont {Le}\ \emph {et~al.}(2013)\citenamefont {Le},
  \citenamefont {Imakaev}, \citenamefont {Mirny},\ and\ \citenamefont
  {Laub.}}]{le}%
  \BibitemOpen
  \bibfield  {author} {\bibinfo {author} {\bibfnamefont {T.~B.~K.}\
  \bibnamefont {Le}}, \bibinfo {author} {\bibfnamefont {M.~V.}\ \bibnamefont
  {Imakaev}}, \bibinfo {author} {\bibfnamefont {L.~A.}\ \bibnamefont {Mirny}},
  \ and\ \bibinfo {author} {\bibfnamefont {M.~T.}\ \bibnamefont {Laub.}},\
  }\href {\doibase 10.1126/science.1242059} {\bibfield  {journal} {\bibinfo
  {journal} {Science}\ }\textbf {\bibinfo {volume} {342}},\ \bibinfo {pages}
  {731} (\bibinfo {year} {2013})}\BibitemShut {NoStop}%
\bibitem [{\citenamefont {Jun}\ and\ \citenamefont {Mulder}(2006)}]{jun}%
  \BibitemOpen
  \bibfield  {author} {\bibinfo {author} {\bibfnamefont {S.}~\bibnamefont
  {Jun}}\ and\ \bibinfo {author} {\bibfnamefont {B.}~\bibnamefont {Mulder}},\
  }\href {\doibase 10.1073/pnas.0605305103} {\bibfield  {journal} {\bibinfo
  {journal} {Proceedings of the National Academy of Sciences}\ }\textbf
  {\bibinfo {volume} {103}},\ \bibinfo {pages} {12388} (\bibinfo {year}
  {2006})},\ \Eprint
  {http://arxiv.org/abs/http://www.pnas.org/content/103/33/12388.full.pdf}
  {http://www.pnas.org/content/103/33/12388.full.pdf} \BibitemShut {NoStop}%
\bibitem [{\citenamefont {Lieberman-Aiden}\ \emph {et~al.}(2009)\citenamefont
  {Lieberman-Aiden}, \citenamefont {van Berkum}, \citenamefont {Williams},
  \citenamefont {Imakaev}, \citenamefont {Ragoczy}, \citenamefont {Telling},
  \citenamefont {Amit}, \citenamefont {Lajoie}, \citenamefont {Sabo},
  \citenamefont {Dorschner}, \citenamefont {Sandstrom}, \citenamefont
  {Bernstein}, \citenamefont {Bender}, \citenamefont {Groudine}, \citenamefont
  {Gnirke}, \citenamefont {Stamatoyannopoulos}, \citenamefont {Mirny},
  \citenamefont {Lander},\ and\ \citenamefont {Dekker}}]{aiden}%
  \BibitemOpen
  \bibfield  {author} {\bibinfo {author} {\bibfnamefont {E.}~\bibnamefont
  {Lieberman-Aiden}}, \bibinfo {author} {\bibfnamefont {N.~L.}\ \bibnamefont
  {van Berkum}}, \bibinfo {author} {\bibfnamefont {L.}~\bibnamefont
  {Williams}}, \bibinfo {author} {\bibfnamefont {M.}~\bibnamefont {Imakaev}},
  \bibinfo {author} {\bibfnamefont {T.}~\bibnamefont {Ragoczy}}, \bibinfo
  {author} {\bibfnamefont {A.}~\bibnamefont {Telling}}, \bibinfo {author}
  {\bibfnamefont {I.}~\bibnamefont {Amit}}, \bibinfo {author} {\bibfnamefont
  {B.~R.}\ \bibnamefont {Lajoie}}, \bibinfo {author} {\bibfnamefont {P.~J.}\
  \bibnamefont {Sabo}}, \bibinfo {author} {\bibfnamefont {M.~O.}\ \bibnamefont
  {Dorschner}}, \bibinfo {author} {\bibfnamefont {R.}~\bibnamefont
  {Sandstrom}}, \bibinfo {author} {\bibfnamefont {B.}~\bibnamefont
  {Bernstein}}, \bibinfo {author} {\bibfnamefont {M.~A.}\ \bibnamefont
  {Bender}}, \bibinfo {author} {\bibfnamefont {M.}~\bibnamefont {Groudine}},
  \bibinfo {author} {\bibfnamefont {A.}~\bibnamefont {Gnirke}}, \bibinfo
  {author} {\bibfnamefont {J.}~\bibnamefont {Stamatoyannopoulos}}, \bibinfo
  {author} {\bibfnamefont {L.~A.}\ \bibnamefont {Mirny}}, \bibinfo {author}
  {\bibfnamefont {E.~S.}\ \bibnamefont {Lander}}, \ and\ \bibinfo {author}
  {\bibfnamefont {J.}~\bibnamefont {Dekker}},\ }\href {\doibase
  10.1126/science.1181369} {\bibfield  {journal} {\bibinfo  {journal}
  {Science}\ }\textbf {\bibinfo {volume} {326}},\ \bibinfo {pages} {289}
  (\bibinfo {year} {2009})}\BibitemShut {NoStop}%
\bibitem [{\citenamefont {Tjong}\ \emph {et~al.}(2012)\citenamefont {Tjong},
  \citenamefont {Gong}, \citenamefont {Chen}, \citenamefont {L.},\ and\
  \citenamefont {Alber}}]{tjong}%
  \BibitemOpen
  \bibfield  {author} {\bibinfo {author} {\bibfnamefont {H.}~\bibnamefont
  {Tjong}}, \bibinfo {author} {\bibfnamefont {K.}~\bibnamefont {Gong}},
  \bibinfo {author} {\bibnamefont {Chen}}, \bibinfo {author} {\bibnamefont
  {L.}}, \ and\ \bibinfo {author} {\bibfnamefont {F.}~\bibnamefont {Alber}},\
  }\href@noop {} {\bibfield  {journal} {\bibinfo  {journal} {Genome Res.}\
  }\textbf {\bibinfo {volume} {22}},\ \bibinfo {pages} {1295} (\bibinfo {year}
  {2012})}\BibitemShut {NoStop}%
\bibitem [{\citenamefont {Joyeux}(2015)}]{joyeux}%
  \BibitemOpen
  \bibfield  {author} {\bibinfo {author} {\bibfnamefont {M.}~\bibnamefont
  {Joyeux}},\ }\href {http://stacks.iop.org/0953-8984/27/i=38/a=383001}
  {\bibfield  {journal} {\bibinfo  {journal} {Journal of Physics: Condensed
  Matter}\ }\textbf {\bibinfo {volume} {27}},\ \bibinfo {pages} {383001}
  (\bibinfo {year} {2015})}\BibitemShut {NoStop}%
\bibitem [{\citenamefont {Gilbert}\ \emph {et~al.}(2017)\citenamefont
  {Gilbert}, \citenamefont {Marenduzzo},\ and\ \citenamefont
  {Davide}}]{marenduzzo}%
  \BibitemOpen
  \bibfield  {author} {\bibinfo {author} {\bibfnamefont {N.}~\bibnamefont
  {Gilbert}}, \bibinfo {author} {\bibnamefont {Marenduzzo}}, \ and\ \bibinfo
  {author} {\bibnamefont {Davide}},\ }\href {\doibase
  10.1007/s10577-017-9551-2} {\bibfield  {journal} {\bibinfo  {journal}
  {Chromosome Research}\ }\textbf {\bibinfo {volume} {25}},\ \bibinfo {pages}
  {1} (\bibinfo {year} {2017})}\BibitemShut {NoStop}%
\bibitem [{\citenamefont {Brocken}\ \emph {et~al.}(2018)\citenamefont
  {Brocken}, \citenamefont {Tark-Dame},\ and\ \citenamefont {Dame}}]{brocken}%
  \BibitemOpen
  \bibfield  {author} {\bibinfo {author} {\bibfnamefont {D.~J.}\ \bibnamefont
  {Brocken}}, \bibinfo {author} {\bibfnamefont {M.}~\bibnamefont {Tark-Dame}},
  \ and\ \bibinfo {author} {\bibfnamefont {R.~T.}\ \bibnamefont {Dame}},\
  }\href {\doibase https://doi.org/10.1016/j.coisb.2018.02.007} {\bibfield
  {journal} {\bibinfo  {journal} {Current Opinion in Systems Biology}\ }\textbf
  {\bibinfo {volume} {8}},\ \bibinfo {pages} {137 } (\bibinfo {year} {2018})},\
  \bibinfo {note} {• Regulatory and metabolic networks • Special Section:
  Single cell and noise}\BibitemShut {NoStop}%
\bibitem [{\citenamefont {Jonathan D.~Halverson}\ and\ \citenamefont
  {Grosberg}(2014)}]{kremer}%
  \BibitemOpen
  \bibfield  {author} {\bibinfo {author} {\bibfnamefont {K.~K.}\ \bibnamefont
  {Jonathan D.~Halverson}, \bibfnamefont {Jan~Smrek}}\ and\ \bibinfo {author}
  {\bibfnamefont {A.~Y.}\ \bibnamefont {Grosberg}},\ }\href@noop {} {\bibfield
  {journal} {\bibinfo  {journal} {Reports on Progress in Physics}\ }\textbf
  {\bibinfo {volume} {77}} (\bibinfo {year} {2014.})}\BibitemShut {NoStop}%
\bibitem [{\citenamefont {Cagliero}\ \emph {et~al.}(2013)\citenamefont
  {Cagliero}, \citenamefont {Grand}, \citenamefont {Jones}, \citenamefont
  {Jin},\ and\ \citenamefont {O’Sullivan}}]{cagliero}%
  \BibitemOpen
  \bibfield  {author} {\bibinfo {author} {\bibfnamefont {C.}~\bibnamefont
  {Cagliero}}, \bibinfo {author} {\bibfnamefont {R.~S.}\ \bibnamefont {Grand}},
  \bibinfo {author} {\bibfnamefont {M.~B.}\ \bibnamefont {Jones}}, \bibinfo
  {author} {\bibfnamefont {D.~J.}\ \bibnamefont {Jin}}, \ and\ \bibinfo
  {author} {\bibfnamefont {J.~M.}\ \bibnamefont {O’Sullivan}},\ }\href
  {\doibase 10.1093/nar/gkt325} {\bibfield  {journal} {\bibinfo  {journal}
  {Nucleic Acids Res.}\ }\textbf {\bibinfo {volume} {41}},\ \bibinfo {pages}
  {6058} (\bibinfo {year} {2013})}\BibitemShut {NoStop}%
\bibitem [{\citenamefont {Agarwal}\ \emph
  {et~al.}(2018{\natexlab{a}})\citenamefont {Agarwal}, \citenamefont
  {Manjunath}, \citenamefont {Habib},\ and\ \citenamefont {Chatterji}}]{epl}%
  \BibitemOpen
  \bibfield  {author} {\bibinfo {author} {\bibfnamefont {T.}~\bibnamefont
  {Agarwal}}, \bibinfo {author} {\bibfnamefont {G.~P.}\ \bibnamefont
  {Manjunath}}, \bibinfo {author} {\bibfnamefont {F.}~\bibnamefont {Habib}}, \
  and\ \bibinfo {author} {\bibfnamefont {A.}~\bibnamefont {Chatterji}},\ }\href
  {http://stacks.iop.org/0295-5075/121/i=1/a=18004} {\bibfield  {journal}
  {\bibinfo  {journal} {EPL (Europhysics Letters)}\ }\textbf {\bibinfo {volume}
  {121}},\ \bibinfo {pages} {18004} (\bibinfo {year}
  {2018}{\natexlab{a}})}\BibitemShut {NoStop}%
\bibitem [{\citenamefont {Agarwal}\ \emph
  {et~al.}(2018{\natexlab{b}})\citenamefont {Agarwal}, \citenamefont
  {Manjunath}, \citenamefont {Habib}, \citenamefont {Vaddavalli},\ and\
  \citenamefont {Chatterji}}]{jpcm}%
  \BibitemOpen
  \bibfield  {author} {\bibinfo {author} {\bibfnamefont {T.}~\bibnamefont
  {Agarwal}}, \bibinfo {author} {\bibfnamefont {G.~P.}\ \bibnamefont
  {Manjunath}}, \bibinfo {author} {\bibfnamefont {F.}~\bibnamefont {Habib}},
  \bibinfo {author} {\bibfnamefont {P.~L.}\ \bibnamefont {Vaddavalli}}, \ and\
  \bibinfo {author} {\bibfnamefont {A.}~\bibnamefont {Chatterji}},\ }\href
  {http://stacks.iop.org/0953-8984/30/i=3/a=034003} {\bibfield  {journal}
  {\bibinfo  {journal} {Journal of Physics: Condensed Matter}\ }\textbf
  {\bibinfo {volume} {30}},\ \bibinfo {pages} {034003} (\bibinfo {year}
  {2018}{\natexlab{b}})}\BibitemShut {NoStop}%
\bibitem [{\citenamefont {Agarwal}\ \emph {et~al.}()\citenamefont {Agarwal},
  \citenamefont {Manjunath}, \citenamefont {Habib},\ and\ \citenamefont
  {Chatterji}}]{paper3}%
  \BibitemOpen
  \bibfield  {author} {\bibinfo {author} {\bibfnamefont {T.}~\bibnamefont
  {Agarwal}}, \bibinfo {author} {\bibfnamefont {G.~P.}\ \bibnamefont
  {Manjunath}}, \bibinfo {author} {\bibfnamefont {F.}~\bibnamefont {Habib}}, \
  and\ \bibinfo {author} {\bibfnamefont {A.}~\bibnamefont {Chatterji}},\
  }\href@noop {} {\bibinfo  {journal} {Part I of this paper}\ }\BibitemShut
  {NoStop}%
\bibitem [{\citenamefont {Shin}\ \emph {et~al.}(2014)\citenamefont {Shin},
  \citenamefont {Cherstvy},\ and\ \citenamefont {Metzler}}]{ralf}%
  \BibitemOpen
\bibfield  {journal} {  }\bibfield  {author} {\bibinfo {author} {\bibfnamefont
  {J.}~\bibnamefont {Shin}}, \bibinfo {author} {\bibfnamefont {A.~G.}\
  \bibnamefont {Cherstvy}}, \ and\ \bibinfo {author} {\bibfnamefont
  {R.}~\bibnamefont {Metzler}},\ }\href
  {http://stacks.iop.org/1367-2630/16/i=5/a=053047} {\bibfield  {journal}
  {\bibinfo  {journal} {New Journal of Physics}\ }\textbf {\bibinfo {volume}
  {16}},\ \bibinfo {pages} {053047} (\bibinfo {year} {2014})}\BibitemShut
  {NoStop}%
\bibitem [{\citenamefont {McGuffee}\ and\ \citenamefont
  {Elcock}(2010)}]{elcock}%
  \BibitemOpen
  \bibfield  {author} {\bibinfo {author} {\bibfnamefont {S.~R.}\ \bibnamefont
  {McGuffee}}\ and\ \bibinfo {author} {\bibfnamefont {A.~H.}\ \bibnamefont
  {Elcock}},\ }\href {\doibase 10.1371/journal.pcbi.1000694} {\bibfield
  {journal} {\bibinfo  {journal} {PLOS Computational Biology}\ }\textbf
  {\bibinfo {volume} {6}},\ \bibinfo {pages} {1} (\bibinfo {year}
  {2010})}\BibitemShut {NoStop}%
\bibitem [{\citenamefont {Kojima}\ \emph {et~al.}(2006)\citenamefont {Kojima},
  \citenamefont {Kubo},\ and\ \citenamefont {Yoshikawa}}]{kojima}%
  \BibitemOpen
  \bibfield  {author} {\bibinfo {author} {\bibfnamefont {M.}~\bibnamefont
  {Kojima}}, \bibinfo {author} {\bibfnamefont {K.}~\bibnamefont {Kubo}}, \ and\
  \bibinfo {author} {\bibfnamefont {K.}~\bibnamefont {Yoshikawa}},\ }\href
  {\doibase 10.1063/1.2145752} {\bibfield  {journal} {\bibinfo  {journal} {The
  Journal of Chemical Physics}\ }\textbf {\bibinfo {volume} {124}},\ \bibinfo
  {pages} {024902} (\bibinfo {year} {2006})}\BibitemShut {NoStop}%
\bibitem [{\citenamefont {Wu}\ \emph {et~al.}(2018)\citenamefont {Wu},
  \citenamefont {Swain}, \citenamefont {Kuijpers}, \citenamefont {Zheng},
  \citenamefont {Felter}, \citenamefont {Guurink}, \citenamefont {Chaudhuri},
  \citenamefont {Mulder},\ and\ \citenamefont {Dekker}}]{wu}%
  \BibitemOpen
  \bibfield  {author} {\bibinfo {author} {\bibfnamefont {F.}~\bibnamefont
  {Wu}}, \bibinfo {author} {\bibfnamefont {P.}~\bibnamefont {Swain}}, \bibinfo
  {author} {\bibfnamefont {L.}~\bibnamefont {Kuijpers}}, \bibinfo {author}
  {\bibfnamefont {X.}~\bibnamefont {Zheng}}, \bibinfo {author} {\bibfnamefont
  {K.}~\bibnamefont {Felter}}, \bibinfo {author} {\bibfnamefont
  {M.}~\bibnamefont {Guurink}}, \bibinfo {author} {\bibfnamefont
  {D.}~\bibnamefont {Chaudhuri}}, \bibinfo {author} {\bibfnamefont
  {B.}~\bibnamefont {Mulder}}, \ and\ \bibinfo {author} {\bibfnamefont
  {C.}~\bibnamefont {Dekker}},\ }\href {\doibase 10.1101/348052} {\  (\bibinfo
  {year} {2018}),\ 10.1101/348052}\BibitemShut {NoStop}%
\bibitem [{\citenamefont {Arnold}\ \emph {et~al.}(2007)\citenamefont {Arnold},
  \citenamefont {Bozorgui}, \citenamefont {Frenkel}, \citenamefont {Ha},\ and\
  \citenamefont {Jun}}]{axel}%
  \BibitemOpen
  \bibfield  {author} {\bibinfo {author} {\bibfnamefont {A.}~\bibnamefont
  {Arnold}}, \bibinfo {author} {\bibfnamefont {B.}~\bibnamefont {Bozorgui}},
  \bibinfo {author} {\bibfnamefont {D.}~\bibnamefont {Frenkel}}, \bibinfo
  {author} {\bibfnamefont {B.-Y.}\ \bibnamefont {Ha}}, \ and\ \bibinfo {author}
  {\bibfnamefont {S.}~\bibnamefont {Jun}},\ }\href {\doibase 10.1063/1.2799513}
  {\bibfield  {journal} {\bibinfo  {journal} {The Journal of Chemical Physics}\
  }\textbf {\bibinfo {volume} {127}},\ \bibinfo {pages} {164903} (\bibinfo
  {year} {2007})}\BibitemShut {NoStop}%
\bibitem [{\citenamefont {Jung}\ \emph {et~al.}(2009)\citenamefont {Jung},
  \citenamefont {Jun},\ and\ \citenamefont {Ha}}]{jung}%
  \BibitemOpen
  \bibfield  {author} {\bibinfo {author} {\bibfnamefont {Y.}~\bibnamefont
  {Jung}}, \bibinfo {author} {\bibfnamefont {S.}~\bibnamefont {Jun}}, \ and\
  \bibinfo {author} {\bibfnamefont {B.-Y.}\ \bibnamefont {Ha}},\ }\href
  {\doibase 10.1103/physreve.79.061912} {\bibfield  {journal} {\bibinfo
  {journal} {Physical Review E}\ }\textbf {\bibinfo {volume} {79}} (\bibinfo
  {year} {2009}),\ 10.1103/physreve.79.061912}\BibitemShut {NoStop}%
\bibitem [{\citenamefont {Kang}\ \emph {et~al.}(2015)\citenamefont {Kang},
  \citenamefont {Yoon}, \citenamefont {Thirumalai},\ and\ \citenamefont
  {Hyeon}}]{hongsuk}%
  \BibitemOpen
  \bibfield  {author} {\bibinfo {author} {\bibfnamefont {H.}~\bibnamefont
  {Kang}}, \bibinfo {author} {\bibfnamefont {Y.-G.}\ \bibnamefont {Yoon}},
  \bibinfo {author} {\bibfnamefont {D.}~\bibnamefont {Thirumalai}}, \ and\
  \bibinfo {author} {\bibfnamefont {C.}~\bibnamefont {Hyeon}},\ }\href
  {\doibase 10.1103/PhysRevLett.115.198102} {\bibfield  {journal} {\bibinfo
  {journal} {Phys. Rev. Lett.}\ }\textbf {\bibinfo {volume} {115}},\ \bibinfo
  {pages} {198102} (\bibinfo {year} {2015})}\BibitemShut {NoStop}%
\bibitem [{\citenamefont {Hacker}\ \emph {et~al.}(2017)\citenamefont {Hacker},
  \citenamefont {Li},\ and\ \citenamefont {Elcock}}]{william}%
  \BibitemOpen
  \bibfield  {author} {\bibinfo {author} {\bibfnamefont {W.~C.}\ \bibnamefont
  {Hacker}}, \bibinfo {author} {\bibfnamefont {S.}~\bibnamefont {Li}}, \ and\
  \bibinfo {author} {\bibfnamefont {A.~H.}\ \bibnamefont {Elcock}},\ }\href
  {\doibase 10.1093/nar/gkx541} {\bibfield  {journal} {\bibinfo  {journal}
  {Nucleic Acid Research}\ }\textbf {\bibinfo {volume} {45}},\ \bibinfo {pages}
  {7541} (\bibinfo {year} {2017})}\BibitemShut {NoStop}%
\bibitem [{\citenamefont {Chaudhuri}\ and\ \citenamefont
  {Mulder}(2012)}]{debashish}%
  \BibitemOpen
  \bibfield  {author} {\bibinfo {author} {\bibfnamefont {D.}~\bibnamefont
  {Chaudhuri}}\ and\ \bibinfo {author} {\bibfnamefont {B.~M.}\ \bibnamefont
  {Mulder}},\ }\href {\doibase 10.1103/PhysRevLett.108.268305} {\bibfield
  {journal} {\bibinfo  {journal} {Phys. Rev. Lett.}\ }\textbf {\bibinfo
  {volume} {108}},\ \bibinfo {pages} {268305} (\bibinfo {year}
  {2012})}\BibitemShut {NoStop}%
\bibitem [{\citenamefont {Ha}\ and\ \citenamefont {Jung}(2015)}]{byha}%
  \BibitemOpen
  \bibfield  {author} {\bibinfo {author} {\bibfnamefont {B.-Y.}\ \bibnamefont
  {Ha}}\ and\ \bibinfo {author} {\bibfnamefont {Y.}~\bibnamefont {Jung}},\
  }\href {\doibase 10.1039/c4sm02734e} {\bibfield  {journal} {\bibinfo
  {journal} {Soft Matter}\ }\textbf {\bibinfo {volume} {11}},\ \bibinfo {pages}
  {2333} (\bibinfo {year} {2015})}\BibitemShut {NoStop}%
\bibitem [{\citenamefont {Reisner}\ \emph {et~al.}(2012)\citenamefont
  {Reisner}, \citenamefont {Pedersen},\ and\ \citenamefont {Austin}}]{austin}%
  \BibitemOpen
  \bibfield  {author} {\bibinfo {author} {\bibfnamefont {W.}~\bibnamefont
  {Reisner}}, \bibinfo {author} {\bibfnamefont {J.~N.}\ \bibnamefont
  {Pedersen}}, \ and\ \bibinfo {author} {\bibfnamefont {R.~H.}\ \bibnamefont
  {Austin}},\ }\href {\doibase 10.1088/0034-4885/75/10/106601} {\bibfield
  {journal} {\bibinfo  {journal} {Reports on Progress in Physics}\ }\textbf
  {\bibinfo {volume} {75}},\ \bibinfo {pages} {106601} (\bibinfo {year}
  {2012})}\BibitemShut {NoStop}%
\bibitem [{\citenamefont {Tung}\ \emph {et~al.}(2015)\citenamefont {Tung},
  \citenamefont {Composto}, \citenamefont {Riggleman},\ and\ \citenamefont
  {Winey}}]{tung}%
  \BibitemOpen
  \bibfield  {author} {\bibinfo {author} {\bibfnamefont {W.-S.}\ \bibnamefont
  {Tung}}, \bibinfo {author} {\bibfnamefont {R.~J.}\ \bibnamefont {Composto}},
  \bibinfo {author} {\bibfnamefont {R.~A.}\ \bibnamefont {Riggleman}}, \ and\
  \bibinfo {author} {\bibfnamefont {K.~I.}\ \bibnamefont {Winey}},\ }\href
  {\doibase 10.1021/acs.macromol.5b00085} {\bibfield  {journal} {\bibinfo
  {journal} {Macromolecules}\ }\textbf {\bibinfo {volume} {48}},\ \bibinfo
  {pages} {2324} (\bibinfo {year} {2015})}\BibitemShut {NoStop}%
\bibitem [{\citenamefont {Wang}\ \emph {et~al.}(2011)\citenamefont {Wang},
  \citenamefont {Tree},\ and\ \citenamefont {Dorfman}}]{wang}%
  \BibitemOpen
  \bibfield  {author} {\bibinfo {author} {\bibfnamefont {Y.}~\bibnamefont
  {Wang}}, \bibinfo {author} {\bibfnamefont {D.~R.}\ \bibnamefont {Tree}}, \
  and\ \bibinfo {author} {\bibfnamefont {K.~D.}\ \bibnamefont {Dorfman}},\
  }\href {\doibase 10.1021/ma201277e} {\bibfield  {journal} {\bibinfo
  {journal} {Macromolecules}\ }\textbf {\bibinfo {volume} {44}},\ \bibinfo
  {pages} {6594} (\bibinfo {year} {2011})}\BibitemShut {NoStop}%
\bibitem [{\citenamefont {Dai}\ \emph {et~al.}(2014)\citenamefont {Dai},
  \citenamefont {van~der Maarel},\ and\ \citenamefont {Doyle}}]{dai}%
  \BibitemOpen
  \bibfield  {author} {\bibinfo {author} {\bibfnamefont {L.}~\bibnamefont
  {Dai}}, \bibinfo {author} {\bibfnamefont {J.}~\bibnamefont {van~der Maarel}},
  \ and\ \bibinfo {author} {\bibfnamefont {P.~S.}\ \bibnamefont {Doyle}},\
  }\href {\doibase 10.1021/ma500326w} {\bibfield  {journal} {\bibinfo
  {journal} {Macromolecules}\ }\textbf {\bibinfo {volume} {47}},\ \bibinfo
  {pages} {2445} (\bibinfo {year} {2014})}\BibitemShut {NoStop}%
\bibitem [{\citenamefont {Milchev}(2011)}]{milchev}%
  \BibitemOpen
  \bibfield  {author} {\bibinfo {author} {\bibfnamefont {A.}~\bibnamefont
  {Milchev}},\ }\href {\doibase 10.1088/0953-8984/23/10/103101} {\bibfield
  {journal} {\bibinfo  {journal} {Journal of Physics: Condensed Matter}\
  }\textbf {\bibinfo {volume} {23}},\ \bibinfo {pages} {103101} (\bibinfo
  {year} {2011})}\BibitemShut {NoStop}%
\bibitem [{eff()}]{effectivecl}%
  \BibitemOpen
  \href@noop {} {\bibinfo  {journal} {Effective CLs: Some of the CLs in BC-CL
  set are not independent of each other. For e.g. suppose monomer indexed i and
  j constitute a CL, then another CL which constitute monomer index i+1 and j+1
  is not independent from the former CL.}\ }\BibitemShut {NoStop}%
\bibitem [{\citenamefont {Chiariello}\ \emph {et~al.}(2016)\citenamefont
  {Chiariello}, \citenamefont {Annunziatella}, \citenamefont {Bianco},
  \citenamefont {Esposito},\ and\ \citenamefont {Nicodemi}}]{nicodemi}%
  \BibitemOpen
\bibfield  {journal} {  }\bibfield  {author} {\bibinfo {author} {\bibfnamefont
  {A.~M.}\ \bibnamefont {Chiariello}}, \bibinfo {author} {\bibfnamefont
  {C.}~\bibnamefont {Annunziatella}}, \bibinfo {author} {\bibfnamefont
  {S.}~\bibnamefont {Bianco}}, \bibinfo {author} {\bibfnamefont
  {A.}~\bibnamefont {Esposito}}, \ and\ \bibinfo {author} {\bibfnamefont
  {M.}~\bibnamefont {Nicodemi}},\ }\href {\doibase 10.1038/srep29775}
  {\bibfield  {journal} {\bibinfo  {journal} {Scientific Reports}\ }\textbf
  {\bibinfo {volume} {6}} (\bibinfo {year} {2016}),\
  10.1038/srep29775}\BibitemShut {NoStop}%
\bibitem [{\citenamefont {Fudenberg}\ \emph {et~al.}(2016)\citenamefont
  {Fudenberg}, \citenamefont {Imakaev}, \citenamefont {Lu}, \citenamefont
  {Goloborodko}, \citenamefont {Abdennur},\ and\ \citenamefont
  {Mirny}}]{geoffrey}%
  \BibitemOpen
  \bibfield  {author} {\bibinfo {author} {\bibfnamefont {G.}~\bibnamefont
  {Fudenberg}}, \bibinfo {author} {\bibfnamefont {M.}~\bibnamefont {Imakaev}},
  \bibinfo {author} {\bibfnamefont {C.}~\bibnamefont {Lu}}, \bibinfo {author}
  {\bibfnamefont {A.}~\bibnamefont {Goloborodko}}, \bibinfo {author}
  {\bibfnamefont {N.}~\bibnamefont {Abdennur}}, \ and\ \bibinfo {author}
  {\bibfnamefont {L.~A.}\ \bibnamefont {Mirny}},\ }\href {\doibase
  10.1016/j.celrep.2016.04.085} {\bibfield  {journal} {\bibinfo  {journal}
  {Cell Reports}\ }\textbf {\bibinfo {volume} {15}},\ \bibinfo {pages} {2038}
  (\bibinfo {year} {2016})}\BibitemShut {NoStop}%
\bibitem [{\citenamefont {Stavans}\ and\ \citenamefont
  {Oppenheim}(2006)}]{amos}%
  \BibitemOpen
  \bibfield  {author} {\bibinfo {author} {\bibfnamefont {J.}~\bibnamefont
  {Stavans}}\ and\ \bibinfo {author} {\bibfnamefont {A.}~\bibnamefont
  {Oppenheim}},\ }\href {http://stacks.iop.org/1478-3975/3/i=4/a=R01}
  {\bibfield  {journal} {\bibinfo  {journal} {Physical Biology}\ }\textbf
  {\bibinfo {volume} {3}},\ \bibinfo {pages} {R1} (\bibinfo {year}
  {2006})}\BibitemShut {NoStop}%
\end{thebibliography}%
\clearpage

%%%%%%%%%%%%%%%%%%%%%%%%%%%%%%%%%%%%%%%%%%%%%%%%%%%%%%
%MERGING Supplementary

\begin{center}
\textbf{\large Supplementary Materials}
\end{center}
\beginsupplement
\section{List of cross-linked monomers in our simulations.}

In the following table, we list the monomers which are cross-linked to model the constraints for
the DNA of bacteria {\em C. crescentus} and {\em E. coli} for the $BC^{'}$ CLs set.

This table has been generated by analysis of raw data obtained from C. Cagliero et. al., Nucleic Acids Res, {\bf 41}, 6058-6071 (2013) and Tung B. et. al., Science, {\bf 342}, 731-734 (2013).

\begin{longtable}{|l|l|l|l|l|}%[!tbh]

%  \centering
%\hline
%\begin{tabular}{|l|l|l|l|l|l|l|l|l|l}
\hline
- & \multicolumn{2}{c|}{$BC^{'}$ ({\em C. crescentus})} &\multicolumn{2}{c|}{$BC^{'}$ ({\em E. coli})} \\
\hline
\midrule
Serial  & Monomer & Monomer & Monomer & Monomer \\
no.  &  index-1 & index-2 &  index-1 & index-2  \\ [2pt]
\hline
\hline \\ 
1 & 1 & 4017 & 1 & 4642 \\ [2pt]
2 & 289 & 1985 & 16 & 2515 \\ [2pt]
3 & 289 & 1986 & 17 & 2516 \\[2pt]
4 & 290 & 1987 & 20 & 1051\\  [2pt]
5 & 290 & 1987 & 21 & 1050 \\[2pt]
6 & 468 & 564 & 21 & 3584 \\[2pt]
7 &  469 & 565 &224 & 2731\\[2pt]
8 & 470 & 566 & 224 & 4209 \\[2pt]
9 & 470 & 567 & 225 & 2730\\[2pt]
10 & 471 & 567 & 225 & 2731  \\[2pt]
11 & 541 & 2494 &226 & 2729\\[2pt]
12 &  541 & 2907 &226 & 2730\\[2pt]
13 &  541 & 2957 &226 & 3428\\[2pt]
14 &  541 & 2958 & 227 & 2728  \\[2pt]
15 &  641 & 683 & 227 & 2729\\[2pt]
16 &  693 & 2875 &228 & 2727\\[2pt]
17 &  693 & 2876 &228 & 2728\\[2pt]
18 &  693 & 2945 &229 & 2727\\[2pt]
19 &  694 & 2876 &229 & 4213\\[2pt]
20 &  710 & 1890 &229 & 4214\\[2pt]
21 &  710 & 3145 &271 &4509\\[2pt]
22 & 733 & 1890 &272 &4508\\[2pt]
23 &  847 & 3912 &274 & 1301\\[2pt]
24 & 1032 & 1437 &275 & 1300\\[2pt]
25 & 1032 & 3580 &280 & 1050\\[2pt]
26 & 1032 & 3850 &280 & 1051 \\[2pt]
27 & 1032 & 3851 &291 & 1051\\[2pt]
28 & 1033 & 3579 &316 & 392\\[2pt]
29 &  1033 & 3850 &316 & 393\\[2pt]
30 &  1057 & 3331 &317 & 392\\[2pt]
31 &  1261 & 2613 &317 & 1095\\[2pt]
32 &  1369 & 2250 &317 & 2172 \\[2pt]
33 &  1370 & 1393 &382 & 1469\\[2pt]
34 &  1370 & 2249 &383 & 1469\\[2pt]
35 &  1370 & 2250 &393 & 567 \\[2pt]
36 &  1393 & 3119 &526 & 1529\\[2pt]
37 &  1398 & 3455 &527 & 1529\\[2pt]
38 &  1399 & 3454 &527 & 1530\\[2pt]
39 &  1399 & 3455 &575 & 1301\\[2pt]
40 &  1421 & 1657 &609 & 2515\\[2pt]
41 &  1437 & 3579 &688 & 1301\\ [2pt]
42 & 1437 & 3580 &730 & 3763\\[2pt]
43 &  2249 & 2803 &731 & 3764 \\[2pt]
44 &  2250 & 2804 &732 & 3765\\[2pt]
45 &  2303 & 3240 &733 & 735\\[2pt]
46 &  2493 & 2907 &733 & 3765\\[2pt]
47 &  2494 & 2906 &733 & 3766\\[2pt]
48 &  2494 & 2906 &735 & 3766\\[2pt]
49 &  2581 & 2583 &1208 & 1210\\[2pt]
50 &  2581 & 2584 &1269 & 1271\\[2pt]
51 &  2582 & 2584 &1301 & 3132\\[2pt]
52 & 2803 & 3119 &1433 & 1635\\[2pt]
53 &  2837 & 3767 &1434 & 1634\\[2pt]
54 &  2838 & 3768 &1435 & 1633\\[2pt]
55 &  2839 & 3769 &1470 & 2998\\[2pt]
56 & 2842 & 3772 &1470 & 3186\\[2pt]
57 &  2875 & 2945 &1470 &4509\\[2pt]
58 & 2875 & 2946 &1533 & 3626\\[2pt]
59 &  3348 & 3377 &1571 & 3667 \\[2pt]
60 & 3579 & 3850 &1572 & 3667\\[2pt]
61 & - & -      & 1572 & 3668 \\[2pt]
62 & - & -      &  2728 & 3945\\[2pt]
63 & - & -      &  2729 & 3945\\[2pt]
64 & - & -      &  2729 & 4038\\[2pt]
65 & - & -      &   2730 & 3943\\[2pt]
66 & - & -      &  3427 & 4038\\[2pt]
67 & - & -      &  3429 & 3942\\[2pt]
68 & - & -      &  3471 & 4177\\[2pt]
69 & - & -      &  3472 & 4176\\[2pt]
70 & - & -      & 3620 & 3763\\[2pt]
71 & - & -      &  3620 & 3764\\[2pt]
72 & - & -      &  3621 & 3764\\[2pt]
73 & - & -      & 3622 & 3765\\[2pt]
74 & - & -      & 3622 & 3766\\[2pt]
75 & - & -      &  3623 & 3766\\[2pt]
76 & - & -      & 3623 & 3768\\[2pt]
77 & - & -      & 3944 & 4210\\[2pt]
\hline
\bottomrule

%\end{tabular}
\caption{\label{tab:table11}
The table shows the list of pair of monomers which constitute the  CLs for {\em E. coli} and {\em C. crescentus} corresponding to the $BC^{'}$ CL sets mentioned in the main paper. These CLs are
used as an input to our simulation by constraining these monomers to be at a distance $a$ from each other.
The first monomer with label $1$ and the last monomer labeled $4642$ (or $4017$) are linked together because the bacterial DNA is a ring polymer.
}
\end{longtable}

\begin{figure}[!hbt]
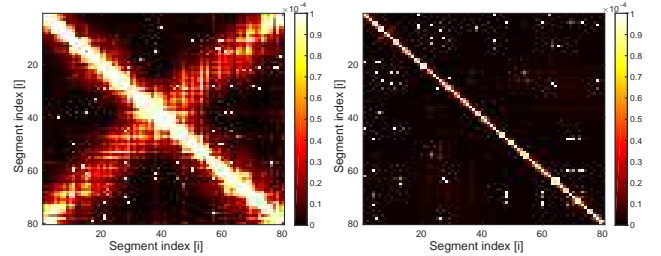

\includegraphics[width=0.48\columnwidth]{matrix_caul_manip.eps}
\includegraphics[width=0.48\columnwidth]{matrix_ecoli_manip.eps}
\caption{\label{fig101}
The figure show the coarse-grained experimental contact maps of bacteria {\em C. crescentus} and {\em E. coli}. The x-axis and y-axis represent the segment index numbered $1,2...80$ and the color represents the probability of the two segments to be found close to each other in Hi-C experiments. The graph has been taken from the supplementary materials of the paper T. Agarwal et. al., Europhysics letters, {\bf 121}, 18004 (2018) with the permission from the publisher.}
\end{figure}

\clearpage

\end{document}